\documentclass[%
reprint,
superscriptaddress,
nofootinbib,
amsmath,amssymb,
aps,
]{revtex4-2}

\usepackage{graphicx}
\usepackage{dcolumn}
\usepackage{bm}
\usepackage{float}
\usepackage{xcolor}
\usepackage[caption=false]{subfig}
\usepackage[hidelinks]{hyperref}

\begin{document}

\preprint{APS/123-QED}

\title{Next-to-leading order perturbative QCD predictions for exclusive $J/\psi$ photoproduction in oxygen-oxygen and lead-lead collisions at the LHC}

\author{K.~J.~Eskola}
\email{kari.eskola@jyu.fi}

\affiliation{University of Jyvaskyla, Department of Physics, P.O. Box 35, FI-40014 University of Jyvaskyla, Finland}%
\affiliation{Helsinki Institute of Physics, P.O. Box 64, FI-00014 University of Helsinki, Finland}

\author{C.~A.~Flett}%
\email{chris.a.flett@jyu.fi}

\affiliation{%
University of Jyvaskyla, Department of Physics, P.O. Box 35, FI-40014 University of Jyvaskyla, Finland}%
\affiliation{Helsinki Institute of Physics, P.O. Box 64, FI-00014 University of Helsinki, Finland}
 
\author{V.~Guzey}%
\email{vadim.a.guzey@jyu.fi}
\affiliation{University of Jyvaskyla, Department of Physics, P.O. Box 35, FI-40014 University of Jyvaskyla, Finland}%
\affiliation{Helsinki Institute of Physics, P.O. Box 64, FI-00014 University of Helsinki, Finland}

\author{T.~L\"oyt\"ainen}
\email{topi.m.o.loytainen@jyu.fi}
 
\affiliation{University of Jyvaskyla, Department of Physics, P.O. Box 35, FI-40014 University of Jyvaskyla, Finland}%
\affiliation{Helsinki Institute of Physics, P.O. Box 64, FI-00014 University of Helsinki, Finland}
 
\author{H.~Paukkunen}
\email{hannu.paukkunen@jyu.fi}
 
\affiliation{University of Jyvaskyla, Department of Physics, P.O. Box 35, FI-40014 University of Jyvaskyla, Finland}%
\affiliation{Helsinki Institute of Physics, P.O. Box 64, FI-00014 University of Helsinki, Finland}

\date{\today}

\begin{abstract}
We present predictions for the cross sections of coherent $J/\psi$ photoproduction in lead-lead and oxygen-oxygen ultraperipheral collisions (UPCs) as a function of the $J/\psi$ rapidity at the LHC in the framework of collinear factorization at next-to-leading order (NLO) in perturbative QCD. Taking generalized parton distribution functions in their forward limit and using the EPPS21, nNNPDF3.0, and nCTEQ15WZSIH nuclear parton distribution functions, we update our recent results for Pb-Pb collisions, make detailed predictions for O-O collisions for several beam energy configurations, and examine the ratio of O-O and Pb-Pb UPC cross sections. We show that the latter observable allows one to significantly reduce the scale uncertainty of NLO predictions for this process.
\end{abstract}

\maketitle

\section{\label{Sec:Intro}Introduction}
Traditionally parton distribution functions (PDFs) and their nuclear counterparts, nuclear PDFs (nPDFs), have been determined from inclusive processes such as lepton-hadron deep inelastic scattering (DIS) and the production of leptons pairs (Drell-Yan process), light and heavy mesons, dijets, and gauge bosons in hadron-hadron scattering, see Refs.~\cite{Gao:2017yyd,Rojo:2019uip,Kovarik:2019xvh,Ethier:2020way} for recent reviews. The determination of proton and nuclear PDFs has become an active branch of phenomenological applications of quantum chromodynamics (QCD), for recent examples of global fits of PDFs and nPDFs, see~\cite{Hou:2019efy,NNPDF:2021njg,PDF4LHCWorkingGroup:2022cjn,Kovarik:2015cma,Eskola:2016oht,Khanpour:2020zyu,Eskola:2021nhw,AbdulKhalek:2022fyi,Helenius:2021tof}. However, despite the dramatic progress in the methodology of PDF extraction from the available data, including an account of higher-order (up to next-to-next-to-leading order, NNLO) perturbative QCD corrections, effects of heavy (charm and bottom) quark masses and small-$x$ resummation and the reliance on sophisticated statistical and computational methods (Bayesian and Hessian error estimates and neural networks), the resulting PDFs and nPDFs still suffer from significant uncertainties.  

As a consequence, there is a continuing interest to explore novel kinematics, processes, and observables, which would allow one to obtain additional constraints on the PDFs. In particular, it has been discussed that the exclusive photoproduction of $J/\psi$ mesons on the proton and nuclear targets in the so-called ultraperipheral collisions (UPCs) allows one to probe the gluon density of the target at small momentum fractions $x \sim 10^{-5}- 10^{-3}$ and resolution scales $\mu^2 \sim 3$ GeV$^2$~\cite{Jones:2013pga,Jones:2015nna,Flett:2019pux,Guzey:2013xba,Guzey:2013qza,Guzey:2020ntc} (photoproduction of other quarkonium states, $\psi^{\prime}$ and $\Upsilon$, has also been considered). This is based on the early observation that in the leading logarithmic approximation, i.e., to the leading order (LO) of perturbative QCD (pQCD), the cross section of this process is directly proportional to the gluon density squared~\cite{Ryskin:1992ui}. However, it was later found that the next-to-leading order (NLO) QCD corrections involving both gluon and quark distributions are very large~\cite{Ivanov:2004vd,Jones:2015nna}, which questions the common interpretation in terms of the gluon density. While several methods to stabilize the NLO results have been proposed~\cite{Jones:2016ldq,Flett:2019pux,Flett:2020duk}, further theoretical and phenomenological studies are still required.

We recently performed a detailed study of the cross section of exclusive photoproduction of $J/\psi$ mesons in Pb-Pb UPCs at the Large Hadron Collider (LHC) as a function of the $J/\psi$ rapidity $y$ in the framework of collinear factorization and NLO pQCD, and confirmed the dramatic role of the NLO effects~\cite{Eskola:2022vpi}. In particular, we found that at central rapidities the cross section is dominated by the quark contribution since the gluon one largely cancels in the sum of the LO and the NLO terms. Additionally, even though the scale dependence of our results turned out to be -- as expected -- rather sizable, we determined an ``optimal scale'' allowing for a simultaneous reasonable description of the available Run~1 and Run~2 LHC data on this process. In addition, we observed that the amplitude for this process is predominantly imaginary in a broad range of rapidities with a small window at forward and backward rapidities, where the real part gives the dominant contribution.
   
The purpose of this work is to extend the analysis of Ref.~\cite{Eskola:2022vpi} by (i) updating our previous results for Pb-Pb collisions with three different state-of-the-art nPDF sets, namely, EPPS21~\cite{Eskola:2021nhw}, nNNPDF3.0~\cite{AbdulKhalek:2022fyi}, and nCTEQ15WZSIH~\cite{Kovarik:2015cma}, (ii) making detailed predictions for the ${\rm O}+{\rm O} \rightarrow {\rm O}+J/\psi+{\rm O}$ rapidity-differential cross section for the planned oxygen run at the LHC~\cite{Citron:2018lsq,Brewer:2021kiv}, and (iii) presenting predictions for the ratio of the $J/\psi$ rapidity distributions in Pb-Pb and O-O UPCs. This allows us not only to better control the theoretical uncertainties associated with the nPDFs, but also to tame (reduce) the scale dependence of our NLO results by considering the ratio of $J/\psi$ production with different nuclear collision systems. For the recent predictions of $J/\psi$ photoproduction in O-O UPCs at the LHC in the color dipole framework, see~\cite{Goncalves:2022ret}.

The rest of the paper is organized as follows. In Sec.~\ref{Sec:TheoFrame}, we recapitulate the framework of NLO pQCD coherent exclusive photoproduction of $J/\psi$ in nucleus-nucleus UPCs, pointing out specific extensions to the oxygen beams. Section~\ref{Sec:Results} contains our results, which include updated predictions for $d\sigma({\rm Pb}+{\rm Pb} \to {\rm Pb}+J/\psi+{\rm Pb})/dy$ with the most recent sets of nPDFs and their comparison to all available LHC data on this process, detailed predictions for $d\sigma({\rm O}+{\rm O} \to {\rm O}+J/\psi+{\rm O})/dy$ for the oxygen run with an analysis of the scale dependence and the decomposition into the imaginary and real parts as well as into the gluon and quark contributions, and, finally, predictions for the ratios of the $J/\psi$ rapidity distributions in O-O and Pb-Pb UPCs with an exhaustive analysis of the scale and energy dependence. We discuss and summarize our findings in Sec.~\ref{Sec:Conclusions}.

\section{\label{Sec:TheoFrame} Coherent $J/\psi$ photoproduction in nucleus-nucleus UPCs in NLO pQCD}

In the equivalent photon approximation the cross section of coherent $J/\psi$ photoproduction in UPCs of nuclei (ions) $A_1$ and $A_2$, as a function of the $J/\psi$ rapidity $y$, reads~\cite{Baltz:2007kq}
\begin{equation}\label{XS_plus_minus}
\begin{split}
    \frac{d\sigma^{A_1 A_2\rightarrow A_1 J/\psi A_2} }{dy} = & \left[k \frac{dN_\gamma^{A_1} (k)}{dk} \right]_{k=k^{+}} \sigma^{\gamma A_2 \rightarrow J/\psi A_2}(W^+)  \\
     +& \left[ k \frac{dN_\gamma^{A_2} (k)}{dk} \right]_{k=k^-} \sigma^{A_1 \gamma \rightarrow A_1 J/\psi}(W^{-}) \,, 
\end{split}
\end{equation}
where $k dN_\gamma^{A} (k)/dk$ is the flux of equivalent quasi-real photons emitted by ions $A_1$ and $A_2$, $k$ is the photon energy and $\sigma^{\gamma A \rightarrow VA}(W)$ is the cross section of coherent (without nuclear breakup) $J/\psi$ photoproduction on a nuclear target with $W$ being the collision energy of the photon-nucleon system. The two terms in Eq.~(\ref{XS_plus_minus}) represent two possibilities to arrive at the same final state corresponding either to ion $A_1$ emitting a photon interacting then with ion $A_2$ or ion $A_2$ being a source of photons interacting with ion $A_1$. We define the positive rapidity $y$ in the direction of the ion $A_1$, from which one obtains that the relation between the photon energies $k^\pm$ and the rapidity $y$ is $k^{\pm} = (M_{J/\psi}/2) e^{\pm y}$, where $k^{+}$ and $k^{-}$ refer to ions $A_1$ (positive longitudinal momentum) and $A_2$ (negative longitudinal momentum), respectively, and $M_{J/\psi}$ is the mass of $J/\psi$. The corresponding photon-nucleon system energies are $W^{+}=\sqrt{ 2M_{J/\psi}e^{y}E_2}$ and $W^{-}=\sqrt{2 M_{J/\psi}e^{- y}E_1}$, where $E_2$ and $E_1$ are the per nucleon energies of beams $A_2$ and $A_1$, respectively. For symmetric UPCs, we have $E_2 = E_1 = \sqrt{s_{NN}}/2$, where $\sqrt{s_{NN}}$ is the nucleon-nucleon center-of-mass system (c.m.s.) energy. The interference between the amplitudes, where the photons are emitted by different nuclei, is important only at very small values of the momentum transfer $t$ (very small values of the $J/\psi$ transverse momentum)~\cite{Klein:1999gv} and hence can be safely neglected in the case of the UPC cross section integrated over $t$ which we consider. 

The flux of equivalent photons emitted by a relativistic ion in UPCs is given by a convolution of the impact parameter dependent photon flux $N_{\gamma}^A(k,\Vec{b})$ and the nuclear suppression factor $\Gamma_{AA} (\Vec{b})$,
\begin{equation} \label{Eq:PhotonFlux}
    k\frac{dN_\gamma^A (k)}{dk} = \int d^2 \Vec{b}\, N_\gamma^A (k,\Vec{b}) \Gamma_{AA} (\Vec{b}) \,,
\end{equation}
where $\vec{b}$ is a two-dimensional impact parameter vector denoting the distance between the centers of colliding nuclei in the 
transverse plane. Furthermore, the impact parameter dependent photon flux $N_\gamma^A (k,\Vec{b})$ of a relativistic nucleus $A$ with $Z$ protons can be readily calculated in QED~\cite{Vidovic:1992ik},
\begin{equation}
N_\gamma^A (k,\Vec{b}) = \frac{Z^2 \alpha_{\rm e.m.}}{\pi^2} \Bigg| \int\limits_0^\infty d k_\perp \frac{k_\perp^2 \tilde{F}_A(k_\perp^2 + k^2 /\gamma_L^2)}{k_\perp^2 + k^2 /\gamma_L^2} J_1 (|\vec{b}| k_\perp) \Bigg|^2 \,,
\end{equation}
where $\alpha_{\rm e.m.}$ is the fine-structure constant, $\gamma_L$ is the nucleus Lorentz factor, $J_1$ is the cylindrical modified Bessel function of the first kind and $\tilde{F}_A (t)$ is the nucleus form factor normalized to one, i.e., $\tilde{F}_A(t)=F_A(t)/A$. The nuclear form factor $F_A (t)$, accompanied with the normalization condition $F_A (0) = A$, is in turn given by the standard Fourier transform of the nuclear density $\rho_A(r)$,
\begin{equation} \label{Eq:FormFactor}
F_A (t) = \int d^3 \boldsymbol{r}\, e^{i \boldsymbol{q} \cdot \boldsymbol{r}} \rho_A (r) \,,
\end{equation}
where $t=-|\boldsymbol{q}|^2$.

The nuclear density is well known from measurements of elastic electron-nucleus scattering and is usually parameterized in the form of two-parameter Fermi model (also called Woods-Saxon model) and three-parameter Fermi model (3pF)~\cite{DeVries:1987atn}. The former is typical for heavy nuclei, for lead see~\cite{Eskola:2022vpi}, and latter is usually employed for medium-heavy nuclei. In particular, in this work we use the 3pF parametrization for oxygen
\begin{equation}\label{eq:rho_O}
    \rho_O (r) = \frac{\rho_0 \left( 1 + w \left( \frac{r}{R_A} \right)^2 \right)}{1+ e^{(r-R_A)/d}} \,,
\end{equation}
with the free parameters determined from nuclear charge-density measurements~\cite{DeVries:1987atn},
\begin{equation}
    d=0.513\text{ fm and } w=-0.051 \,.
\end{equation}
For lead we use $d = 0.546$~fm and $w=0$. The effective nuclear radii are here taken from the following empirical parametrization (see e.g., \cite{Helenius:2012wd})\footnote{For the oxygen case this means that the radius, as given in Eq.~(\ref{Eq:NucleusRadius}), is taken in the approximation $w=0$ with the same parameter values as for lead.}
\begin{equation} \label{Eq:NucleusRadius}
R_A/\text{fm} = 1.12 \, A^{1/3} - 0.86 \, A^{-1/3} \,.
\end{equation}

Further, in Eq.~(\ref{Eq:PhotonFlux}), the nuclear suppression factor $\Gamma_{AA}(\Vec{b})$ represents the probability of having no hadronic interactions at impact parameter $\vec{b}$, which can be evaluated using the Glauber model for nucleus-nucleus scattering 
\begin{equation}
\Gamma_{AA}(\Vec{b}) = \exp [-\sigma_{NN}(s_{NN}) T_{AA} (\Vec{b})] \, ,
\label{eq:Gamma_AA}
\end{equation}
where $\sigma_{NN} (s_{NN}) $ is the total nucleon-nucleon cross section~\cite{ParticleDataGroup:2016lqr} and $T_{AA}(\vec{b})=\int d^2 \vec{b^{\prime}} T_A(\vec{b^{\prime}}) T_A(\vec{b}-\vec{b^{\prime}})$ is the nuclear overlap function with $T_{A}(\vec{b})=\int dz \rho_A(\vec{r})$. In Eq.~(\ref{Eq:PhotonFlux}), the effect of $\Gamma_{AA} (b) $ is to suppress the contribution of the $|\vec{b}| < 2 R_A$ region.

For the cross section of coherent $J/\psi$ photoproduction on nuclei $A$ we use the form where the $t$-dependence governed by the nuclear form factor squared $|F_A(t)|^2$, is factorized from the cross section of $J/\psi$ production on bound nucleons $N$ of the nuclear target, i.e., $d\sigma_A^{\gamma N \rightarrow J/\psi N}/dt$ (this is indicated by the subscript),
\begin{equation}
\sigma^{\gamma A \rightarrow J/\psi A} (W) = \frac{d\sigma_A^{\gamma N \rightarrow J/\psi N}}{dt} \bigg|_{t=0} \int\limits_{|t_{\rm min}|}^{\infty} dt^{\prime} |F_A (-t^{\prime})|^2 \,,
\label{eq:cs_gammaA}
\end{equation}
where $|t_{\rm min}|=[M_{J/\psi}^2/(4 k \gamma_L)]^2$. In the case of the $t$-integrated cross section, the factorized form of Eq.~(\ref{eq:cs_gammaA}) approximates with a several-percent precision a more accurate expression that takes into account the correlation between $t$ and $x$, i.e., the correlation between the momentum transfer and the magnitude of nuclear effects (nuclear shadowing) affecting $d\sigma_A^{\gamma N \rightarrow J/\psi N}/dt$, see the discussion in~\cite{Guzey:2016qwo,ALICE:2021tyx}.

The QCD dynamics of the process is contained in the $d\sigma_A^{\gamma N \rightarrow J/\psi N}/dt(t=0)$ cross section. In the framework of collinear factorization for exclusive processes in NLO perturbative QCD and using the non-relativistic (static) approximation for the charmonium wave function, the cross section reads (see~\cite{Eskola:2022vpi,Ivanov:2004vd} for details and references)
\begin{equation}
\frac{d\sigma_A^{\gamma N \rightarrow J/\psi N}}{dt} \bigg|_{t=0}=\frac{1}{W^4} \frac{ 4 \pi^2 \alpha_{\rm e.m.} e_c^2 \langle O_1 \rangle_V}{9 \xi^2 m_c^3}|I(\xi,t=0)|^2 \,,
\label{eq:cs1}
\end{equation}
where the reduced scattering amplitude is given by a convolution of the gluon and quark hard scattering coefficient functions $T_g(x,\xi ,\mu_R , \mu_F)$ and $T_q(x,\xi ,\mu_R , \mu_F)$ with the corresponding gluon and quark generalized parton distribution functions (GPDs) $F^g(x,\xi,t,\mu_F)$ and $F^{q,S}(x,\xi,t,\mu_F)$ of the bound nucleons,
\begin{eqnarray}
I(\xi,t=0)&=&\int^{1}_{-1} dx [T_g(x,\xi ,\mu_R , \mu_F) F^g(x,\xi,t=0,\mu_F) \nonumber\\
&+&T_q(x,\xi ,\mu_R , \mu_F) F^{q,S}(x,\xi,t=0,\mu_F)] \,.
\label{eq:cs2}
\end{eqnarray}
In Eq.~(\ref{eq:cs1}), $e_c=2/3$ and $m_c=M_{J/\psi}/2$ is the charm quark mass in the non-relativistic limit, $\langle O_1 \rangle_V$ is the non-relativistic QCD matrix element associated with the $J/\psi \to l^{+} l^{-}$ leptonic decay, $\xi=\zeta/(2-\zeta)$ is the so-called skewness parameter with $\zeta=M_{J/\psi}^2/W^2$ being an analog of Bjorken $x$ in inclusive DIS. Note that the quark contribution in Eq.~(\ref{eq:cs2}) contains a singlet combination of quark GPDs of four active flavors, $F^{q,S}(x,\xi,t=0,\mu_F)=\sum_{q=u,d,s,c} F^{q}(x,\xi,t=0,\mu_F)$. In our analysis we take the factorization and renormalization scales to be equal, i.e. $\mu=\mu_F=\mu_R$, which sets the term $\sim \beta_0 \ln (\mu_R^2 / \mu_F^2)$ in the NLO gluon contribution to zero, see Eq.~(3.72) in~\cite{Ivanov:2004vd}. We quantify the dependence of our results on the scale choice by varying the scale in the $m_c \leq \mu \leq 2m_c$ interval.

In general, GPDs are complicated nonperturbative distributions depending on two light-cone momentum fractions $x$ and $\xi$ and the momentum transfer $t$ as well as the factorization scale $\mu_F$. However, in the high-energy limit the skewness parameter is very small ($\xi \ll 1$) and its effect on GPDs is expected to be rather moderate. In particular, in the calculation of the $d\sigma_A^{\gamma N \rightarrow J/\psi N}/dt(t=0)$ cross section, theoretical uncertainties associated with detailed modeling of GPDs are expected to be much smaller than the scale and nPDF uncertainties. Therefore, as a first approximation, we neglect the skewness effect and take GPDs in the forward limit, where they can be identified with the usual PDFs,
\begin{equation}
\begin{split}
    &F^g(x,\xi,t=0,\mu_F) = F^g(-x,\xi,t=0,\mu_F) \to xg_A(x,\mu_F) \,, \\
&F^q(x,\xi,t=0,\mu_F)  \to   q_A(x,\mu_F) \,,\\
&F^q(-x,\xi,t=0,\mu_F)  \to  -\bar{q}_A(x,\mu_F) \,,
\end{split}
\label{eq:cs3}
\end{equation}
where now $0 \leq x \leq 1$. The distributions $g_A(x,\mu_F)$, $q_A(x,\mu_F)$ and $\bar{q}_A(x,\mu_F)$ are the usual gluon, quark, and antiquark  nPDFs per nucleon. In our analysis, we use the recent nPDF sets EPPS21~\cite{Eskola:2021nhw}, nNNPDF3.0~\cite{AbdulKhalek:2022fyi} and nCTEQ15WZSIH~\cite{Kusina:2020lyz}.

\section{\label{Sec:Results}Results}
A plan is moving forward that an oxygen-oxygen (O-O) run would be performed at the LHC in Run 3~\cite{Citron:2018lsq,Brewer:2021tyv,Brewer:2021kiv}. In addition to shedding light on the soft QCD dynamics and studying hard scattering with small nuclear systems, it might help to address open questions relating to forward scattering physics. From the point of view of UPC studies, as we pointed out in the Introduction, an analysis of coherent $J/\psi$ photoproduction in O-O UPCs and a comparison to the case of Pb-Pb UPCs can be used to constrain theoretical uncertainties of our NLO pQCD predictions for this process. At the time of writing of this paper, it is not yet completely clear at which nucleon-nucleon c.m.s. energy $\sqrt{s_{NN}}$ the O-O run will be completed. Therefore, we will consider four scenarios with $\sqrt{s_{NN}}=2.76$, 5.02, 6.37~\cite{Brewer:2021tyv}, and 7~TeV~\cite{Citron:2018lsq}, which will help us to better understand the energy dependence of our results. 

\subsection{Photon fluxes and nuclear form factors}

\begin{figure*}
    \centering
    \includegraphics[width=1.0\textwidth]{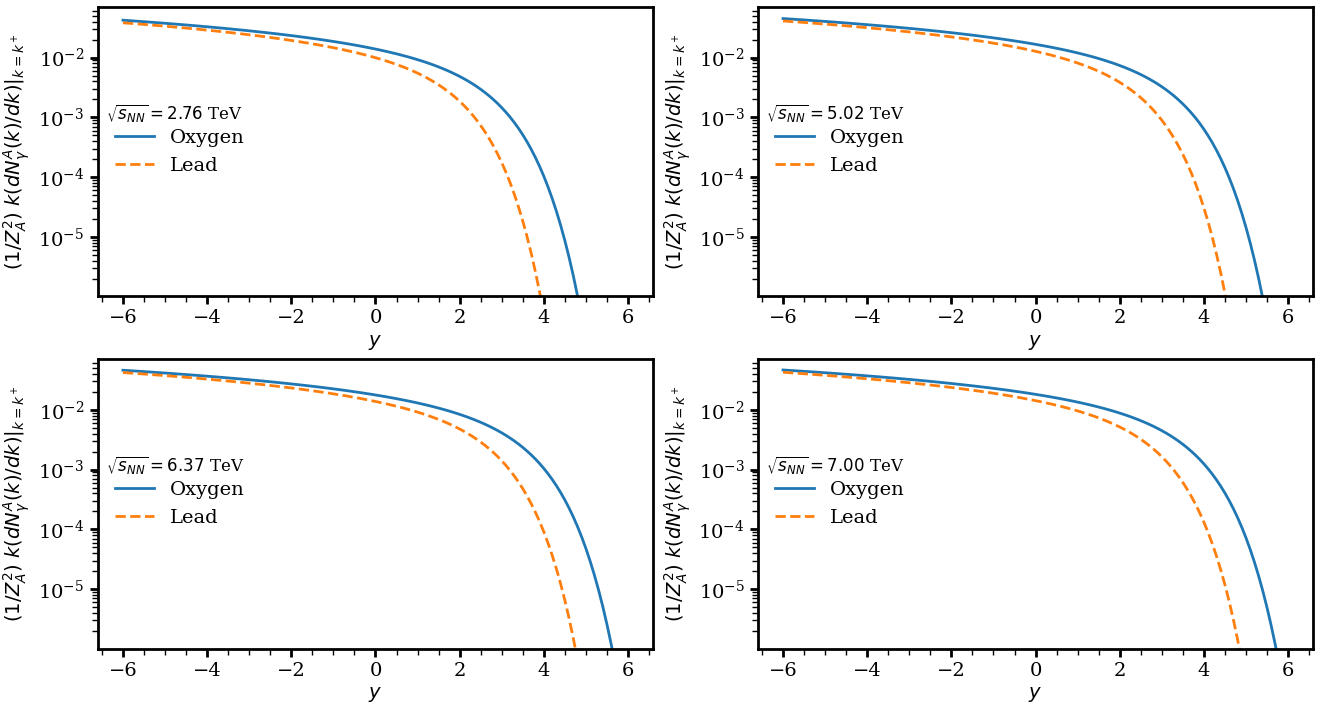}
    \caption{The scaled photon flux $(1/Z^2) kdN_\gamma^A (k)/dk$ as a function of the rapidity $y$ in the plus direction for the oxygen and lead beams for four different values of the c.m.s. energy $\sqrt{s_{NN}}=2.76$, 5.02, 6.37 and 7 TeV.}
    \label{fig:Flux-OxPb}
\end{figure*}

The results for the $ kdN_\gamma^A (k)/dk$ photon flux obtained through Eq.~(\ref{Eq:PhotonFlux}), where $k = (M_{J/\psi}/2) e^{y}$, i.e. positive rapidity $y$ corresponds to large photon energies $k$, for oxygen and lead beams for four different values of the invariant collision energy $\sqrt{s_{NN}}$ are presented in Fig.~\ref{fig:Flux-OxPb}. In the figure, the blue solid curves correspond to the oxygen case and the orange dashed curves to the lead case. In order to conveniently compare the two cases, we normalized the fluxes by the factor of $1/Z^2$ with $Z_{\rm O}=8$ for oxygen and $Z_{\rm Pb} = 82$ for lead. 

One can see from the figure that at negative rapidities (small photon momenta) the photon flux of the lead beam is much larger than that for the oxygen beam, $[kdN_\gamma^{\rm Pb} (k)/dk]/[kdN_\gamma^{\rm O} (k)/dk] \approx Z_{\rm Pb}^2/Z_{\rm O}^2 \approx 100$. At the same time, since the effective nuclear radius of lead is almost 3 times as large as that of oxygen, the spectrum of equivalent photons of lead falls off more rapidly when $y$ is increased (corresponding to an increase in $k$) than that of oxygen. Eventually, for large values of rapidity $y \geq 4.4$ corresponding to $k \geq 120$ GeV, the photon flux for oxygen becomes bigger than that of lead.

\begin{figure*}
    \centering
    \includegraphics[width=1.0\textwidth]{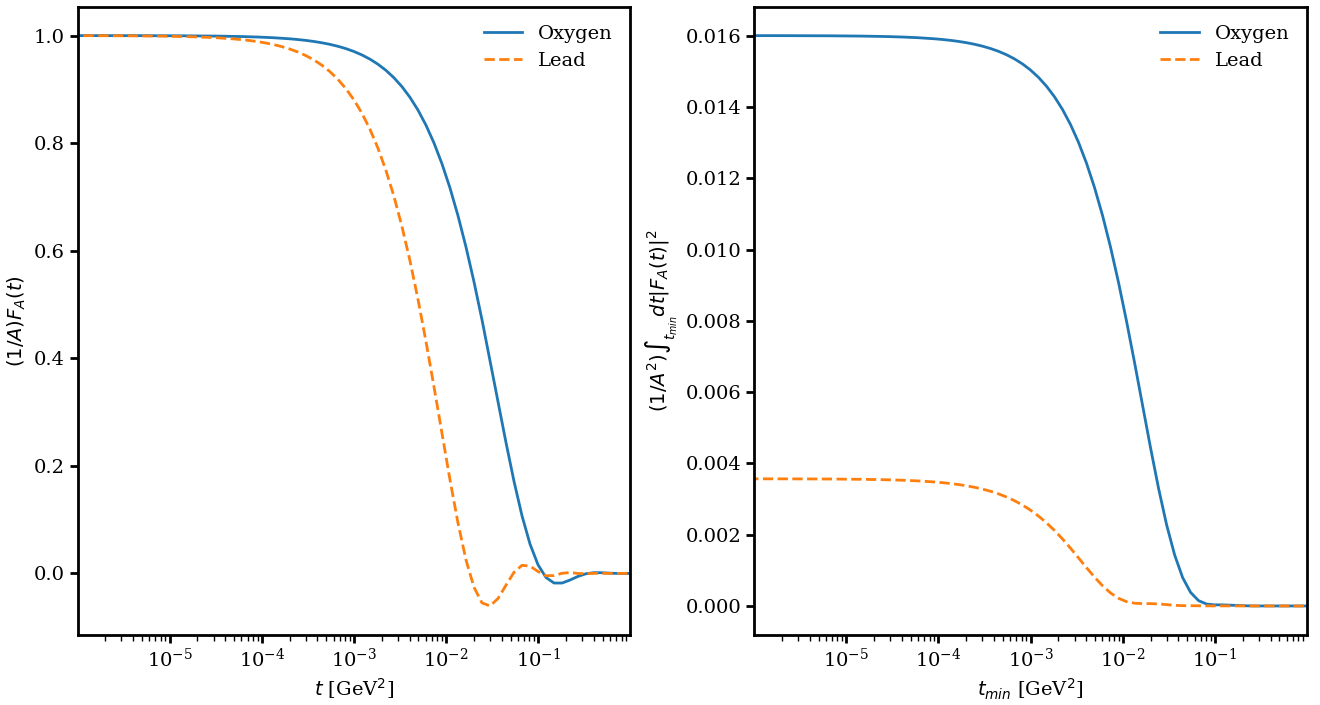}
    \caption{\textbf{Left panel:} Form factors scaled by the corresponding mass number $A$ for oxygen and lead as a function of $t$. The scaled oxygen form factor reaches out further in $y$ than the corresponding lead one. 
    \textbf{Right panel:} The values of the integral of the absolute value of the form factor squared scaled by the square of the corresponding mass number $A$ given as a function of the lower limit $t_{\rm min}$ of the integral. At small enough values of $t_{\rm min}$, the ratio between the oxygen and the lead results is about 4.6. }
    \label{fig:FF-OxPb}
\end{figure*}

We have numerically checked that setting $w=0$ in Eq.~(\ref{eq:rho_O}), i.e., assuming the two-parameter Fermi (2pF) model for oxygen with the same $d$ and $R_A$ parameters, leads to a relative difference of under four percent in the photon flux for the photon energies up to $k \approx 50$~GeV corresponding to the $J/\psi$ rapidities $|y| \leq  3.5$. In addition, we have checked that the photon flux is not sensitive to the exact value of $\sigma_{NN}$ used in $\Gamma_{AA}(\vec{b})$, see Eq.~(\ref{eq:Gamma_AA}). For example, at $\sqrt{s_{NN}}=6.37$ and 7 TeV, calculations for the photon flux with $\sigma_{NN}=95$~mb and $\sigma_{NN}=100$~mb differ by less than 1\% all the way up to $|y|=4$. Thus, we conclude that the major difference between the scaled photon fluxes of the oxygen and lead ions originates from the different effective radii $R_A$ of these nuclei. 

In the left panel of Fig.~\ref{fig:FF-OxPb} we show the results for the oxygen and lead form factors scaled by the corresponding mass number $A$ as they are given by Eq.~(\ref{Eq:FormFactor}). The values of the scaled form factors approach one due to our normalization constraint $F_A (0) = A$ but, as we move to higher values of $t$, we see that the scaled oxygen form factor is the dominant one except for the oscillations at very high values of $t$. But again, since $A_{\rm Pb} \gg A_{\rm O}$, the absolute magnitude of the form factor of lead is the bigger one. Then, in the photoproduction cross section $\sigma^{\gamma A \rightarrow V A}$, we have an integral over the square of the absolute value of the form factor. The right panel of Fig.~\ref{fig:FF-OxPb} shows the values of this integral scaled by the square of the mass number $A$ for both the oxygen (solid blue) and the lead (dashed orange) cases. Similarly to the photon flux, this ratio gets intertwined in the ratios of the cross sections, but at central rapidities $y = 0$ corresponding to $t_{\rm min} \approx 10^{-6}$, we should expect to see a factor of 4.6 from the ratio of the integrals.

\subsection{Rapidity dependent cross sections in Pb-Pb UPCs and comparison with the LHC data}
Based on the NLO pQCD theoretical framework outlined in Sec.~\ref{Sec:TheoFrame}, we present below our predictions for the $d\sigma({\rm Pb}+{\rm Pb} \to {\rm Pb}+J/\psi+{\rm Pb})/dy$ cross section of coherent $J/\psi$ photoproduction in Pb-Pb UPCs as a function of the $J/\psi$ rapidity $y$ at $\sqrt{s_{NN}}=2.76$~TeV (Run~1) and $\sqrt{s_{ NN}}=5.02$~TeV (Run~2) at the LHC and compare them with all available LHC data on this process. We performed our calculations using the most recent EPPS21~\cite{Eskola:2021nhw}, nNNPDF3.0~\cite{AbdulKhalek:2022fyi}, and nCTEQ15WZSIH~\cite{Kusina:2020lyz} sets of nPDFs, which updates our predictions in~\cite{Eskola:2022vpi}.

\begin{figure*}
\centering
    \includegraphics[width=0.49\textwidth]{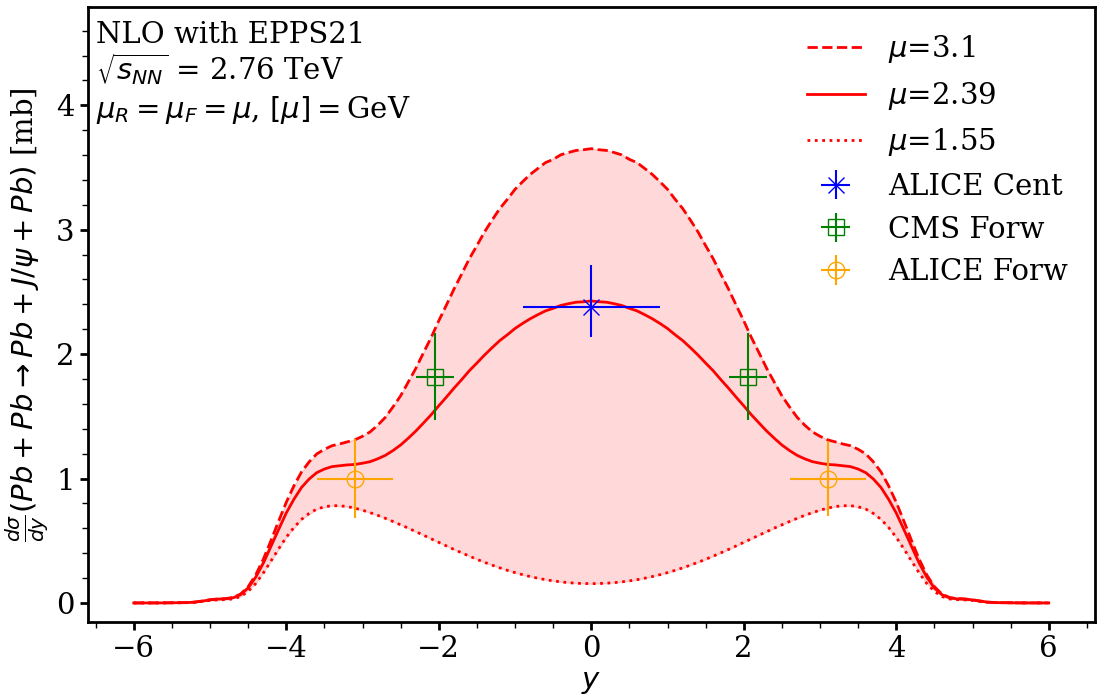}
    \includegraphics[width=0.49\textwidth]{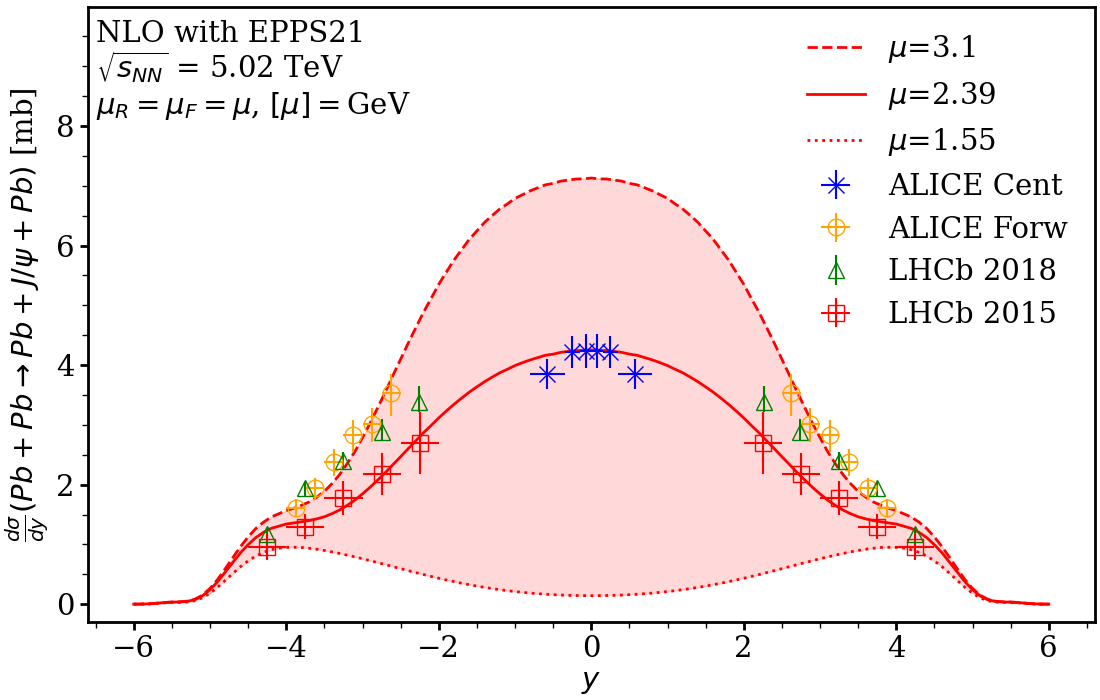}
    \includegraphics[width=0.49\textwidth]{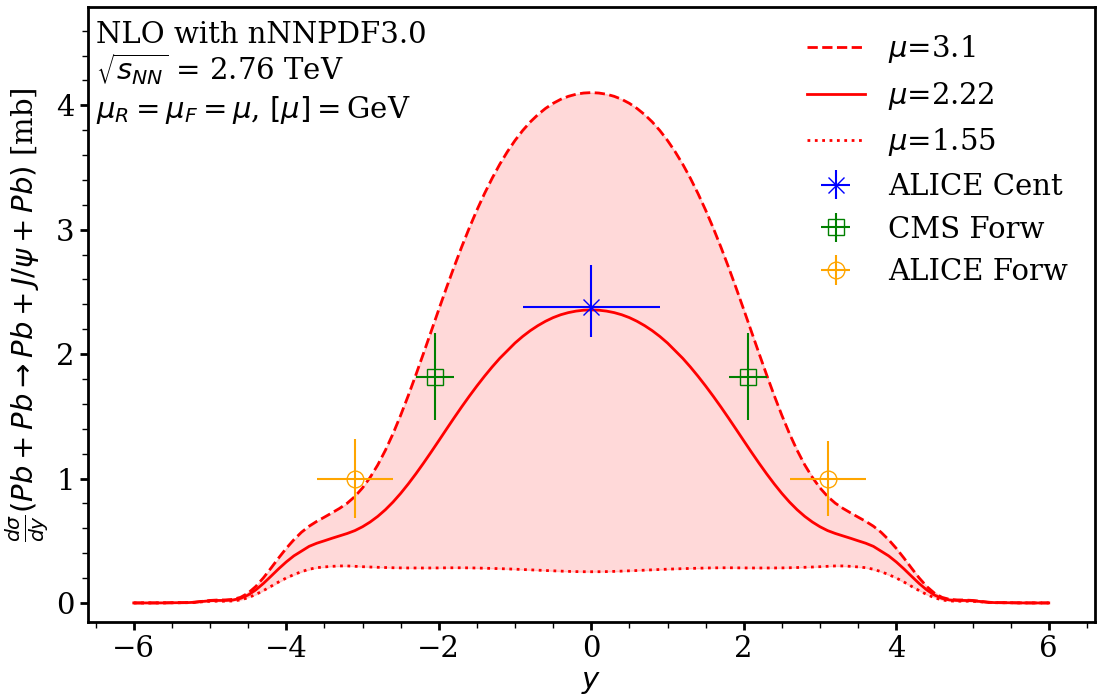}
    \includegraphics[width=0.49\textwidth]{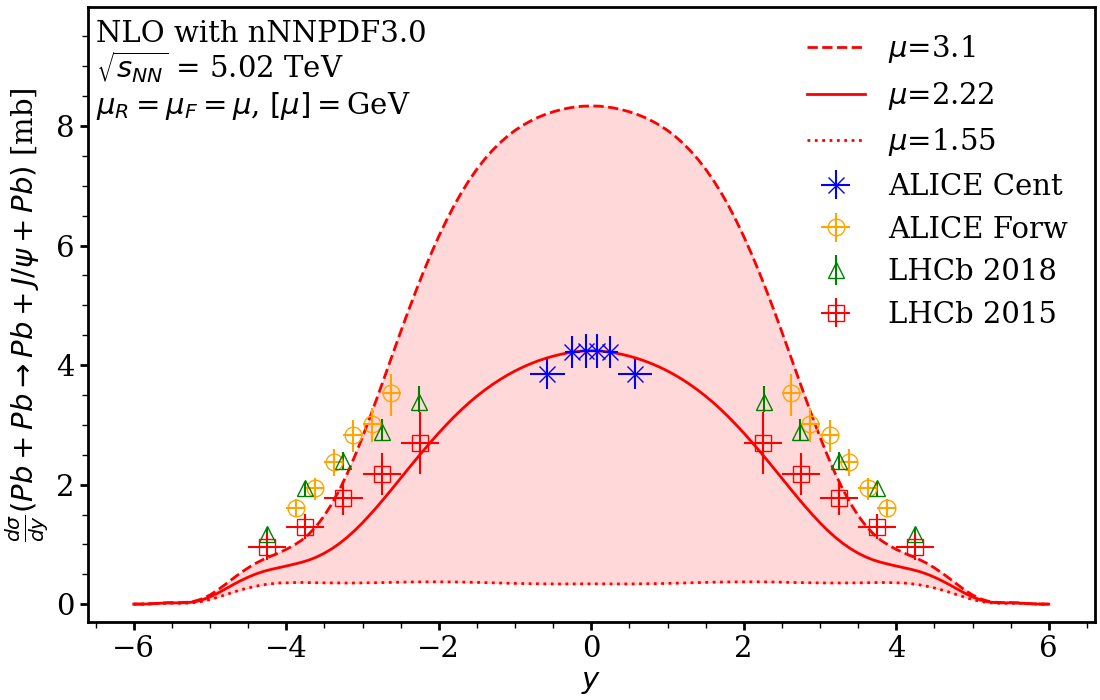}
    \includegraphics[width=0.49\textwidth]{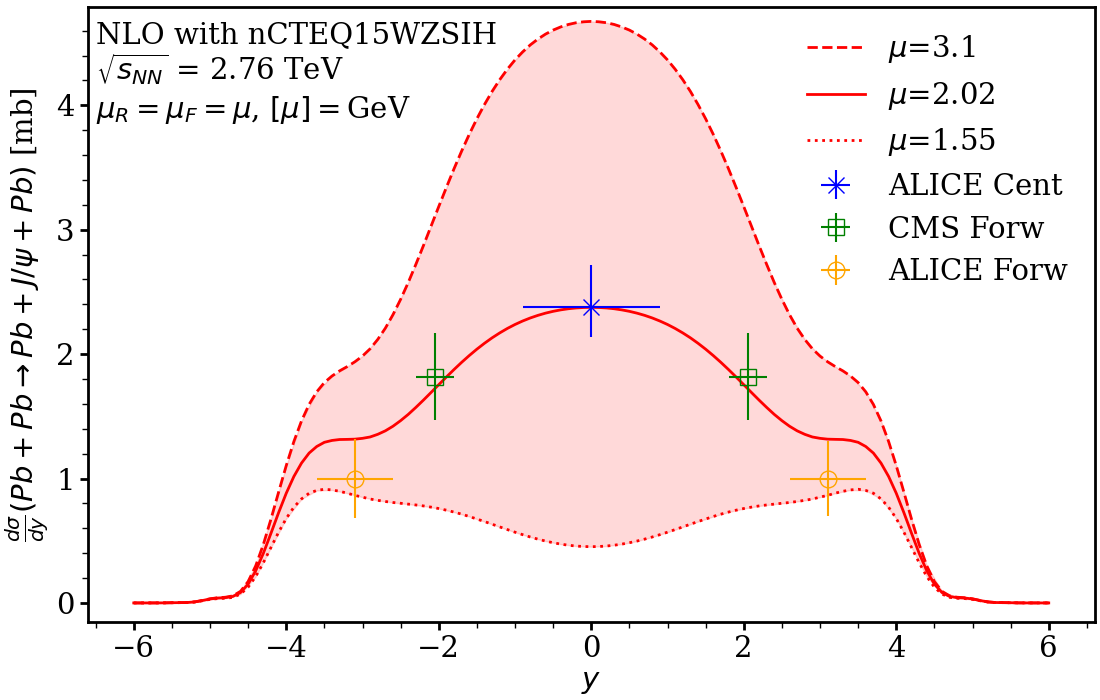}
    \includegraphics[width=0.49\textwidth]{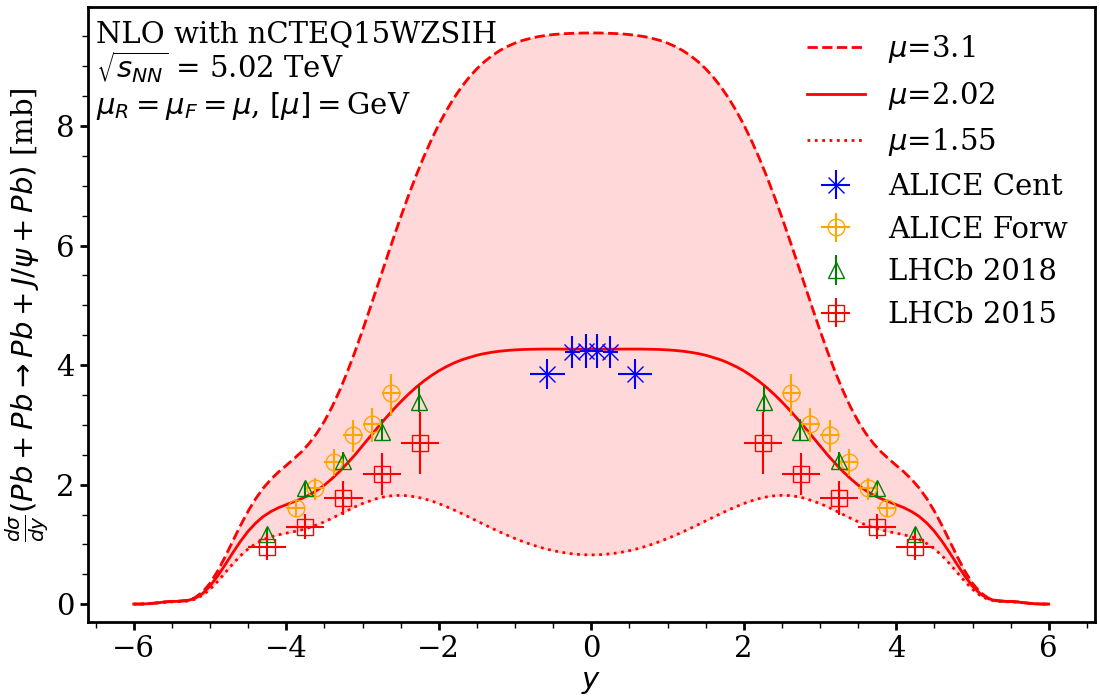}
\caption{The scale dependence of the NLO pQCD predictions for the $d\sigma({\rm Pb}+{\rm Pb} \to {\rm Pb}+J/\psi+{\rm Pb})/dy$ cross section as a function of the rapidity $y$ for Run~1 ($\sqrt{s_{NN}}=2.76$ TeV, left column) and Run~2 ($\sqrt{s_{NN}}=5.02$ TeV, right column) at the LHC and a comparison with the corresponding Run~1~\cite{ALICE:2013wjo,ALICE:2012yye,CMS:2016itn} and Run~2~\cite{ALICE:2021gpt,ALICE:2019tqa,LHCb:2021bfl,Wang:2022iln} data, the statistical and systematic errors added in quadrature. The data have been mirrored with respect to $y=0$. The scale-dependence envelope spans the results corresponding to $\mu=3.1$ GeV (upper dashed curve) and $\mu=1.55$ GeV (lower dotted curve); the solid curve corresponds to the optimal scale. The three rows of panels correspond to EPPS21 (upper), nNNPDF3.0 (middle), and nCTEQ15WZSIH (lower) nPDFs.}
    \label{fig:UpdatedScaleEnvelope}
\end{figure*}

Figure~\ref{fig:UpdatedScaleEnvelope} demonstrates the variation of our predictions due to the choice of the scale $\mu$, which is allowed to vary in the $m_c \leq \mu \leq 2 m_c$ interval ($m_c=M_{J/\psi}/2=1.55$~GeV in the nonrelativistic limit that we use): the upper dashed curves correspond to $\mu=3.1$~GeV, while the lower dotted curves are for $\mu=1.55$~GeV. The solid curve in each panel corresponds to an ``optimal scale", which is chosen to simultaneously describe the central rapidity data available from Run~1 (left panels) and Run~2 (right panels) at the LHC. The Run~1 data at $\sqrt{s_{NN}}=2.76$~TeV include the ALICE data at the central rapidity $|y| < 0.9$~\cite{ALICE:2013wjo} (labeled ``ALICE Cent") and at the forward rapidity $2.6 < |y| < 3.6$~\cite{ALICE:2012yye} labeled ``ALICE Forw") as well as the CMS data in the rapidity interval $1.8 < |y| < 2.3$~\cite{CMS:2016itn} (labeled ``CMS Forw"). The Run~2 data taken at $\sqrt{s_{NN}}=5.02$~TeV are the ALICE data at midrapidity $|y| < 0.8$~\cite{ALICE:2021gpt} (labeled ``ALICE Cent"), the ALICE data at forward rapidities $2.5 < |y| < 4$~\cite{ALICE:2019tqa} (labeled ``ALICE Forw"), the LHCb data at forward rapidities $2 < |y| < 4.5$~\cite{LHCb:2021bfl} (labeled ``LHCb 2015") and their recent update~\cite{Wang:2022iln} (labeled ``LHCb 2018"). The three rows of panels correspond to the results of our calculations using three different sets of nPDFs: EPPS21~\cite{Eskola:2021nhw} (upper row), nNNPDF3.0~\cite{AbdulKhalek:2022fyi} (middle row), and nCTEQ15WZSIH~\cite{Kusina:2020lyz} (lower row). Our analysis shows that the resulting optimal scales $\mu$ slightly differ for different nPDF sets: $\mu=2.39$~GeV for EPPS21, $\mu=2.22$~GeV for nNNPDF3.0, and $\mu=2.02$~GeV for nCTEQ15WZSIH.

A comparison of the results presented in Fig.~\ref{fig:UpdatedScaleEnvelope} with our results in~\cite{Eskola:2022vpi} shows that the difference between our calculations using EPPS21 and EPPS16 is very small with a very similar value of the optimal scale and the same shape of the $y$ dependence as well as the matching magnitude of the scale dependence and the quality of the data description. To be exact, at central rapidity $y=0$, for Run~1 there is a factor of about 22 between the highest scale and the lowest scale results and for Run~2 energy this factor is about 55.

The improvement, when moving from nNNPDF2.0~\cite{AbdulKhalek:2020yuc} (Fig.~10 of~\cite{Eskola:2022vpi}) to the newer nNNPDF3.0 set, is rather dramatic. We find that the shape of the $d\sigma({\rm Pb}+{\rm Pb} \to {\rm Pb}+J/\psi+{\rm Pb})/dy$ cross section at the optimal scale $\mu = 2.22$~GeV is qualitatively similar to that obtained with EPPS16 or EPPS21. Simultaneously, however, the correspondence with the data is slightly worse: while the data at $y \approx 0$ is reproduced by construction, the solid curve somewhat underestimates the data at $|y| \neq 0$. Note that the good agreement with the data at $y \approx 0$ is important for the comparison of the Pb-Pb and O-O UPC data; see the discussion in Sec.~\ref{Subsec:RatiosOfCrossSections}. The ratio between the highest scale and the lowest scale at central rapidity is about 17 for Run~1 and about 26 for Run~2.

In contrast to EPPS21 and nNNPDF3.0, we find that the newest nCTEQ15WZSIH nPDF set actually does better on all accounts. The scale dependence at central rapidity is only about a factor of 10 for Run~1 and about a factor of 12 for Run~2. The curve corresponding to the optimal scale of $\mu = 2.02$ GeV goes through the central rapidity data points in addition to the forward/backward data both at Run~1 and Run~2 energies. Moreover, the nCTEQ15WZSIH nPDF set favors the ALICE forward data~\cite{ALICE:2019tqa} and the newer LHCb 2018 data~\cite{Wang:2022iln} over the 2015 LHCb data~\cite{LHCb:2021bfl}. We have checked that the better agreement of our calculations using the nCTEQ15WZSIH nPDF set with the UPC data is due to the very strongly enhanced strange quark distributions, see Fig.~4 in Ref.~\cite{Duwentaster:2022kpv}. Thus, this process may give an interesting opportunity to obtain new constraints on the elusive strange quark distribution in the proton and nuclei.

For all three sets, when considering the full range of scales $\mu \in [m_c , M_{J/\psi}]$, the scale uncertainty decreases slightly – as was with the earlier EPPS16 set – as we move further away from the central rapidity towards backward and forward rapidities. This is partly because at very large values of rapidity, i.e. $|y|>3$, the photoproduction amplitude receives a large contribution from the $W$-component corresponding to small values of $k$, which in turn means that we are probing the underlying GPDs at high values of $x$, where the scale dependence is constrained rather well. In any case, it is interesting to notice that this rapidity dependence seems to be a common property for both the old and the new nPDF sets (see Fig.~4 in~\cite{Eskola:2022vpi}).

To estimate the PDF uncertainty of our predictions due to the EPPS21 and nCTEQ15WZSIH nPDFs, we used 
the following asymmetric form for the uncertainty $\delta \mathcal{O}^{\pm}$~\cite{Eskola:2016oht}
\begin{equation} \label{Eq:AsymUncert}
    \delta \mathcal{O}^{\pm} = \sqrt{\sum\limits_i \left[^\text{\rm max}_{\rm min} \lbrace \mathcal{O} (S_i^+)- \mathcal{O} (S_0), \mathcal{O}(S^-_i) - \mathcal{O} (S_0),0 \rbrace \right]^2} \,,
\end{equation}
where $\mathcal{O} (S_0)$ denotes the predictions with the central set for the observable $\mathcal{O}$ and $\mathcal{O} (S_i^{\pm})$ correspond to the values calculated with the plus and minus PDF error sets. In the case of nNNPDF3.0, we used the 90\% confidence level (CL) interval prescription~\cite{AbdulKhalek:2022fyi}. All PDF uncertainty calculations are performed at the corresponding values of the optimal scale $\mu$.

\begin{figure*}
    \centering
\includegraphics[width=0.8\textwidth]{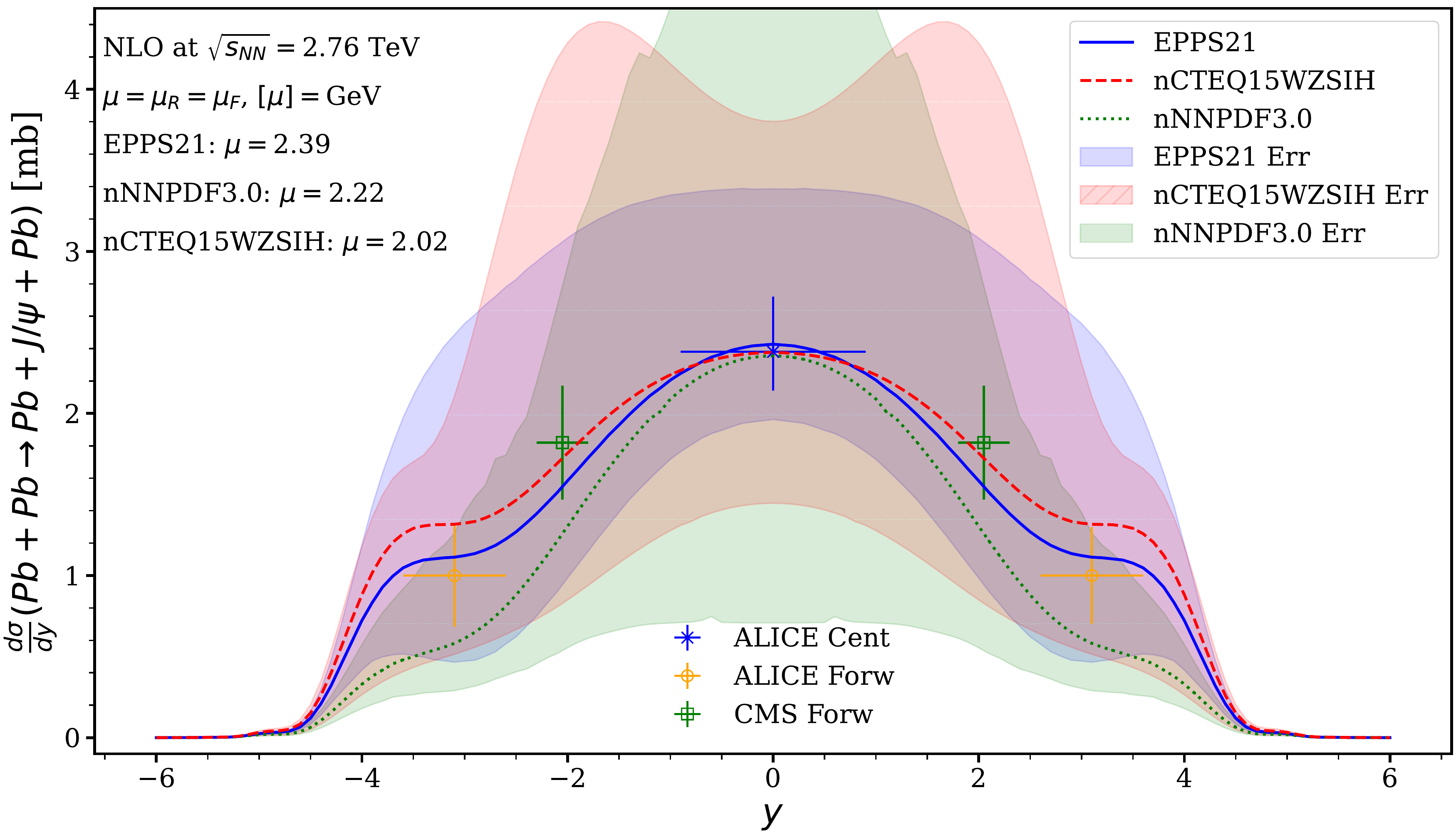}
\includegraphics[width=0.8\textwidth]{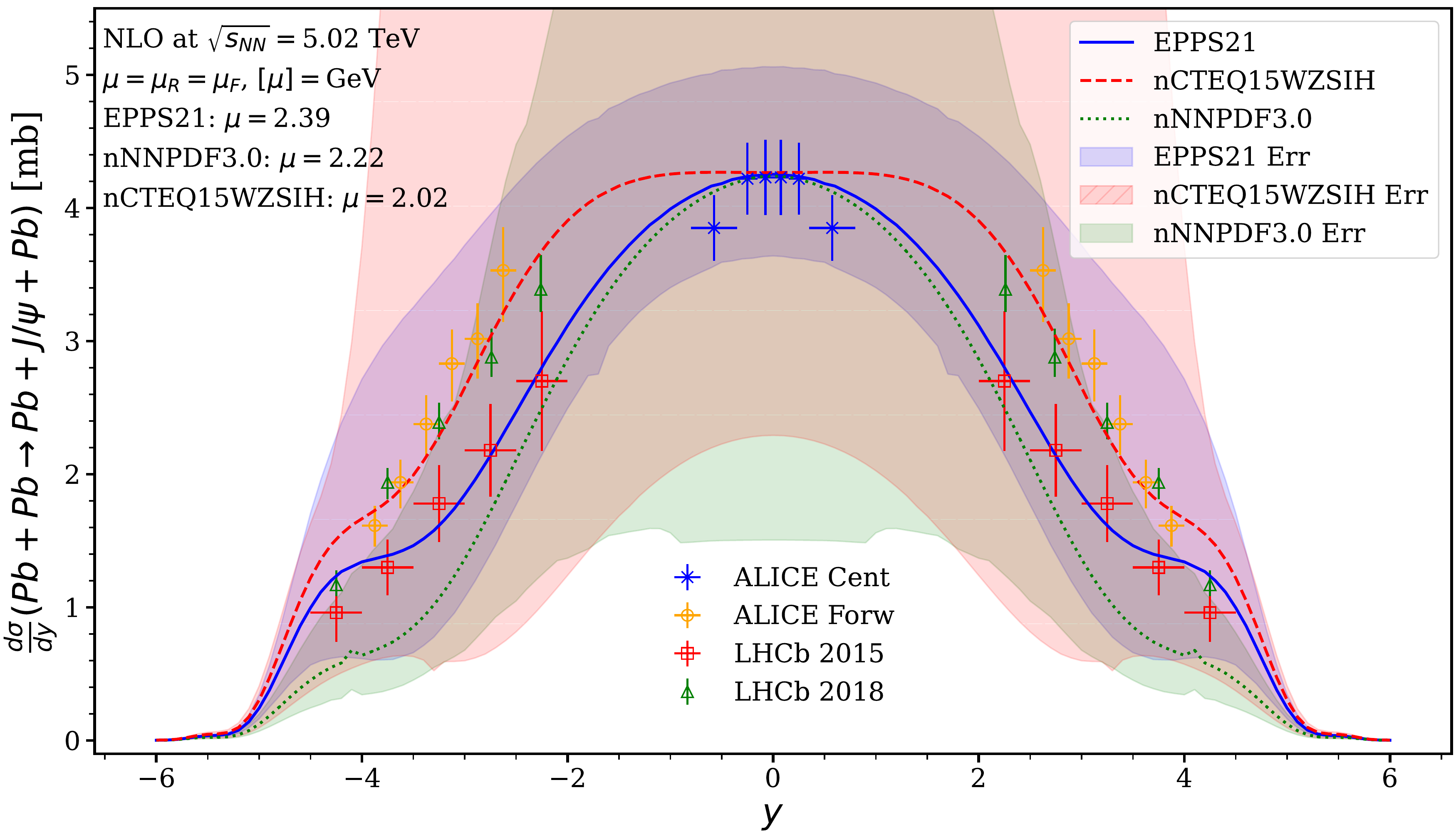}
\caption{The PDF uncertainties of the NLO pQCD predictions for the $d\sigma({\rm Pb}+{\rm Pb} \to {\rm Pb}+J/\psi+{\rm Pb})/dy$ cross section as a function of $y$ for Run~1 (upper) and Run~2 (lower) at the LHC, and a comparison with the corresponding Run~1~\cite{ALICE:2013wjo,ALICE:2012yye,CMS:2016itn} and Run~2~\cite{ALICE:2021gpt,ALICE:2019tqa,LHCb:2021bfl,Wang:2022iln} data, mirrored with respect to $y=0$ and with the statistical and systematic errors added in quadrature. The results corresponding to the central sets of nPDFs are shown by the blue solid (EPPS21), red dashed (nCTEQ15WZSIH), and green dotted (nNNPDF3.0) curves, respectively, and the error bands are represented by the corresponding shaded regions. All calculations are performed at the indicated values of the optimal scale $\mu$.}
    \label{fig:PbPb-PDFuncerts}
\end{figure*}

Figure~\ref{fig:PbPb-PDFuncerts} illustrates the uncertainty of our predictions for the $d\sigma({ \rm Pb}+{\rm Pb} \to {\rm Pb}+J/\psi+{\rm Pb})/dy$ cross section due to errors of nPDFs and compares it with the Run~1 (upper panel) and Run~2 (lower panel) LHC data. The calculations using the central sets of nPDFs are given by the blue solid (EPPS21), red dashed (nCTEQ15WZSIH), and green dotted (nNNPDF3.0) curves and the error bands are represented by the corresponding shaded regions. One can see from the figure that within the PDF uncertainties the framework of NLO pQCD describes the data rather well; the agreement with the data is very good at central rapidity for all three nPDF sets (by construction), continues to be good for nCTEQ15WZSIH in the entire range of measured $y$, but becomes somewhat worse at higher $|y|$ for EPPS21 and NNPDF3.0.

A comparison of our EPPS21 results with the previous EPP16 ones~\cite{Eskola:2022vpi} shows that the full PDF uncertainty band, which receives contributions from varying the parameters of nPDFs and the baseline free proton PDFs, has come down to the order of few millibarns. As we discussed in Ref.~\cite{Eskola:2022vpi}, the free proton CT14nlo PDFs accompanying the EPPS16 nPDFs contain a particular error set dramatically growing at small $x$, which results in an abnormally large small-$x$ uncertainty. In the new EPPS21 nPDFs, where the nuclear effects are correlated with the baseline CT18ANLO~\cite{Hou:2019efy} free proton PDF error sets, this behaviour no longer persists.

One can clearly see from Fig.~\ref{fig:PbPb-PDFuncerts} that the EPPS21 nPDFs correspond to significantly smaller uncertainties than nNNPDF3.0 and CTEQ15WZSIH. In particular, the nNNPDF3.0 uncertainties, which also account for the free proton PDF errors, at central rapidity rise up to around 5.6~mb at Run~1 and up to around 9.5~mb at Run~2. The nCTEQ15WZSIH uncertainties, which account for the nPDF errors only, are smaller both for Run~1 and Run~2 at central rapidities than at $y \approx \pm 2.0$, which leads to a valley-like structure. For instance, for Run~2, the uncertainty rises up to around 8.5~mb at $y = 0$ and then to its maximum of approximately 18~mb at $y \approx \pm 2.5$. As with EPPS21, no single error PDFs set stands out in the nCTEQ15WZSIH and nNNPDF3.0 parametrizations, but the larger uncertainty bands are simply the result of a wider distribution in the underlying error sets.

\subsection{Predictions for rapidity dependent cross sections in O-O UPCs at the LHC}

\begin{figure*}[t]
    \centering
    \includegraphics[width=1.0\textwidth]{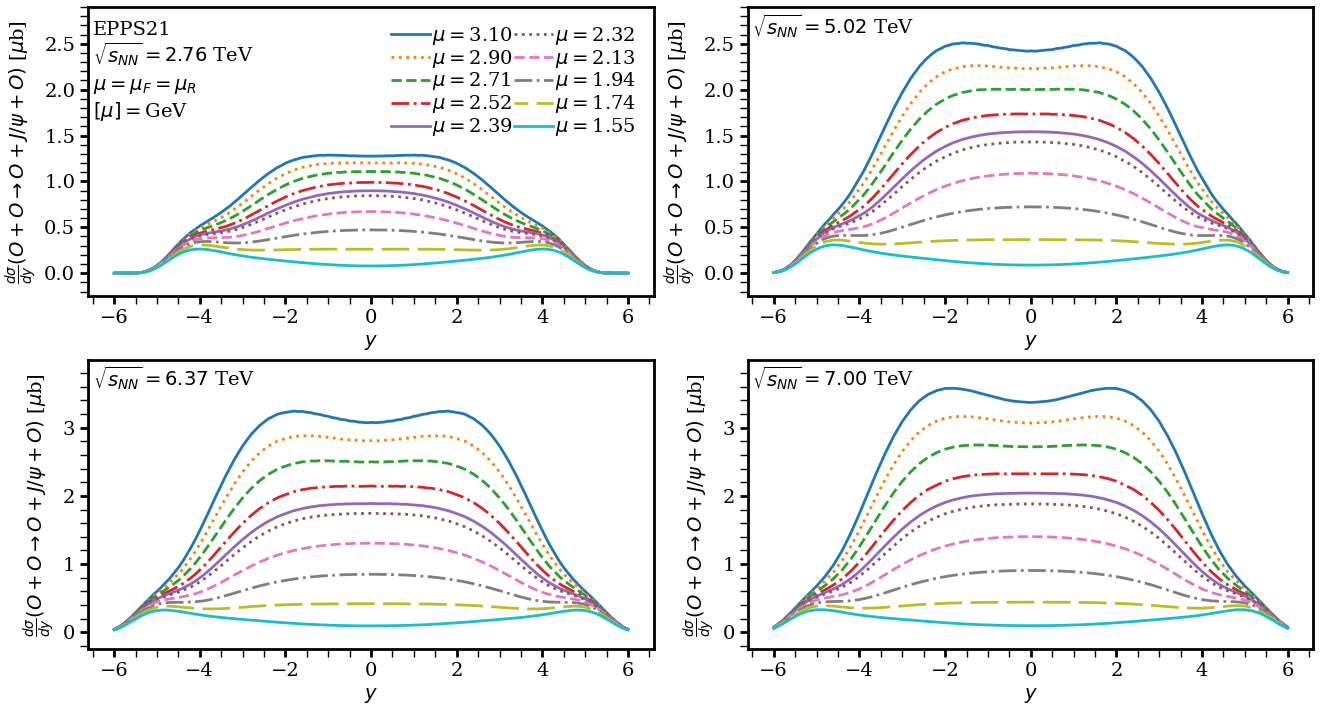}
    \caption{The NLO pQCD results for the rapidity differential cross section of coherent $J/\psi$ photoproduction in O-O UPCs as a function of the rapidity $y$, obtained with the EPPS21 nPDFs at $\sqrt{s_{NN}}=2.76$, 5.02, 6.37 and 7 TeV. The different lines show the results for ten choices of the scale $\mu$ ranging from $\mu = m_c$ (lowest curve) to $\mu = M_{J/\psi}$ (highest curve) with a step of $m_c/8$. The $\mu = 2.39$~GeV ``optimal scale" prediction lies in the middle of this scale-uncertainty envelope.}
    \label{fig:OO_Scales}
\end{figure*}

In this section, we present detailed predictions for the $d\sigma({\rm O}+{\rm O} \to {\rm O}+J/\psi+{\rm O})/dy$ cross section of coherent $J/\psi$ photoproduction in NLO perturbative QCD as a function of the $J/\psi$ rapidity $y$ in oxygen-oxygen UPCs at the LHC. As mentioned above, since the exact energy of O-O collisions is not yet determined, we consider four scenarios with $\sqrt{s_{NN}}=2.76$, 5.02, 6.37 and 7 TeV. In addition to studying the energy dependence of our predictions, this choice enables a direct comparison to the case of Pb-Pb UPCs at $\sqrt{s_{NN}}=2.76$ TeV (Run~1) and $\sqrt{s_{NN}}=5.02$ TeV (Run~2), see the discussion in Sec.~\ref{Subsec:RatiosOfCrossSections}.

Figure~\ref{fig:OO_Scales} illustrates the scale dependence of our predictions and shows our results for $d\sigma({ \rm O}+{ \rm O} \to { \rm O}+J/\psi+{ \rm O})/dy$ with the EPPS21 nPDFs at ten different values of the scale $\mu$ ranging from $\mu=m_c$ up to $\mu=M_{J/\psi}$ for the four different values of $\sqrt{s_{ NN}}$. One can see from the figure that the O-O UPC cross section is approximately 1,000 times smaller than that in the Pb-Pb case primarily due to the much smaller photon flux. On the other hand, the shape of the $y$ dependence is similar in the O-O and Pb-Pb cases: it is rather broad at midrapidity with sloping ``shoulders'' at forward and backward rapidities; higher scales correspond to larger $d\sigma({ \rm O}+{ \rm O} \to { \rm O}+J/\psi+{ \rm O})/dy$, which also tend to develop a valley-like structure at the highest scales of $\mu \approx M_{J/\psi}$.

To quantify the magnitude of the scale dependence, we consider the ratio between the $\mu = M_{J/\psi}$ and $\mu = m_c$ results at $y=0$ which we denote by $R_{\rm scale}$. One can see from Fig.~\ref{fig:OO_Scales} that $R_{\rm scale}$ is of the same order of magnitude as in Pb-Pb collisions starting at $R_{\rm scale} \approx 16$ at $\sqrt{s_{ NN}}=2.76$~TeV and rising up to $R_{\rm scale} \approx 35$ at $\sqrt{s_{NN}}=7$~TeV. We have checked that with nCTEQ15WZSIH the scale dependence is of the same order as with EPPS21: $R_{\rm scale} \approx 12$ at $\sqrt{s_{ NN}}=2.76$~TeV and increasing to approximately $R_{\rm scale}=20$ at $\sqrt{s_{ NN}}=7$~TeV. At the same time, for nNNPDF3.0 the scale dependence is considerably stronger: $R_{\rm scale} \approx 800$ at $\sqrt{s_{ NN}}=2.76$~TeV and increases up to $R_{\rm scale} \approx 2700$ at $\sqrt{s_{ NN}}=7$~TeV. This huge scale dependence is due to the nearly perfect cancellation between the LO and the NLO contributions in both the real and the imaginary parts of the amplitude at the lowest scale of $\mu=m_c$. At forward and backward rapidities over the full range $\mu \in [m_c, M_{J/\psi}]$, the scale dependence is not as strong for all three nPDF sets under consideration.

\begin{figure*}[!ht]
    \centering
    \includegraphics[width=0.93\textwidth]{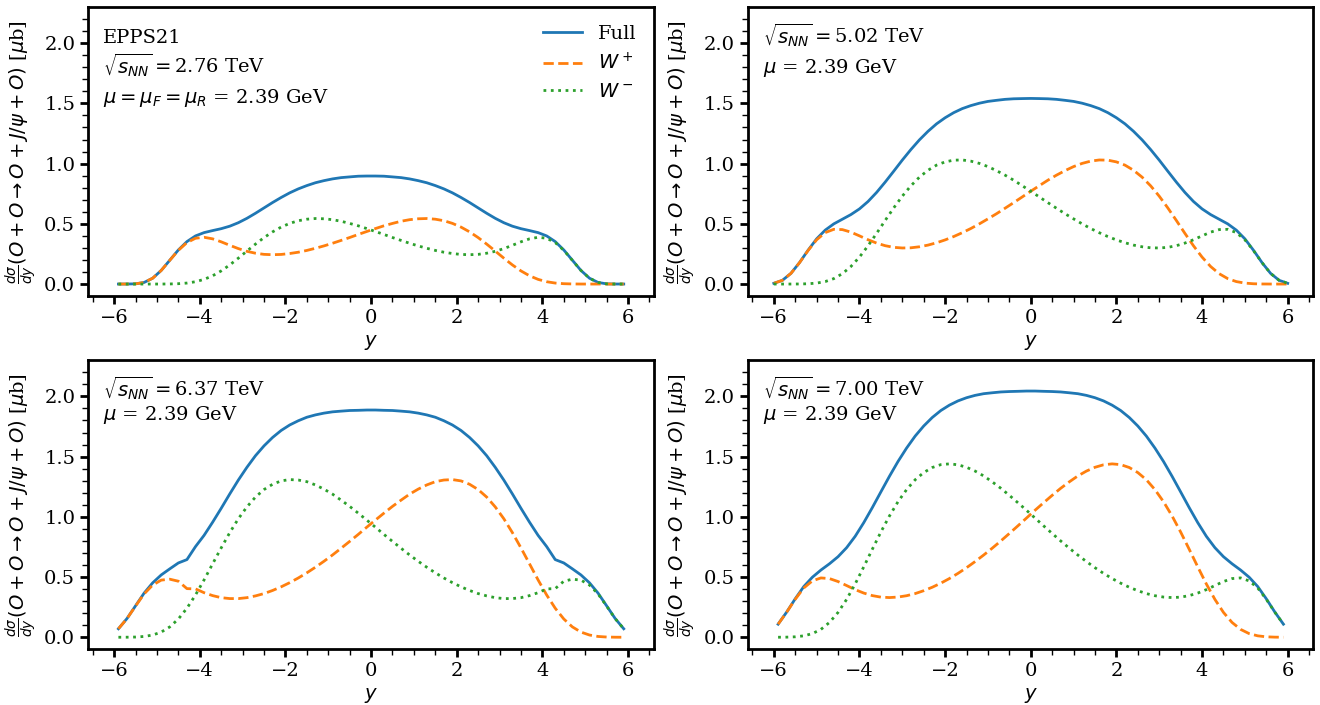}\vspace{-0.3cm}
    \caption{Separation of the NLO pQCD predictions for the $d\sigma({\rm O}+{\rm O} \to {\rm O}+J/\psi+{\rm O})/dy$ cross section of coherent $J/\psi$ photoproduction in O-O UPCs as a function of the rapidity $y$ into the $W^+$ (dashed orange curve) and $W^-$ (dotted green curve) components; the solid blue line is their sum. The calculation employs the EPPS21 nPDFs at $\mu = 2.39$~GeV. The different panels correspond to $\sqrt{s_{ NN}}=2.76$, 5.02, 6.37 and 7 TeV.}
    \label{fig:OO-Wcomp}
\end{figure*}

\begin{figure*}[!ht]
    \centering
    \includegraphics[width=0.93\textwidth]{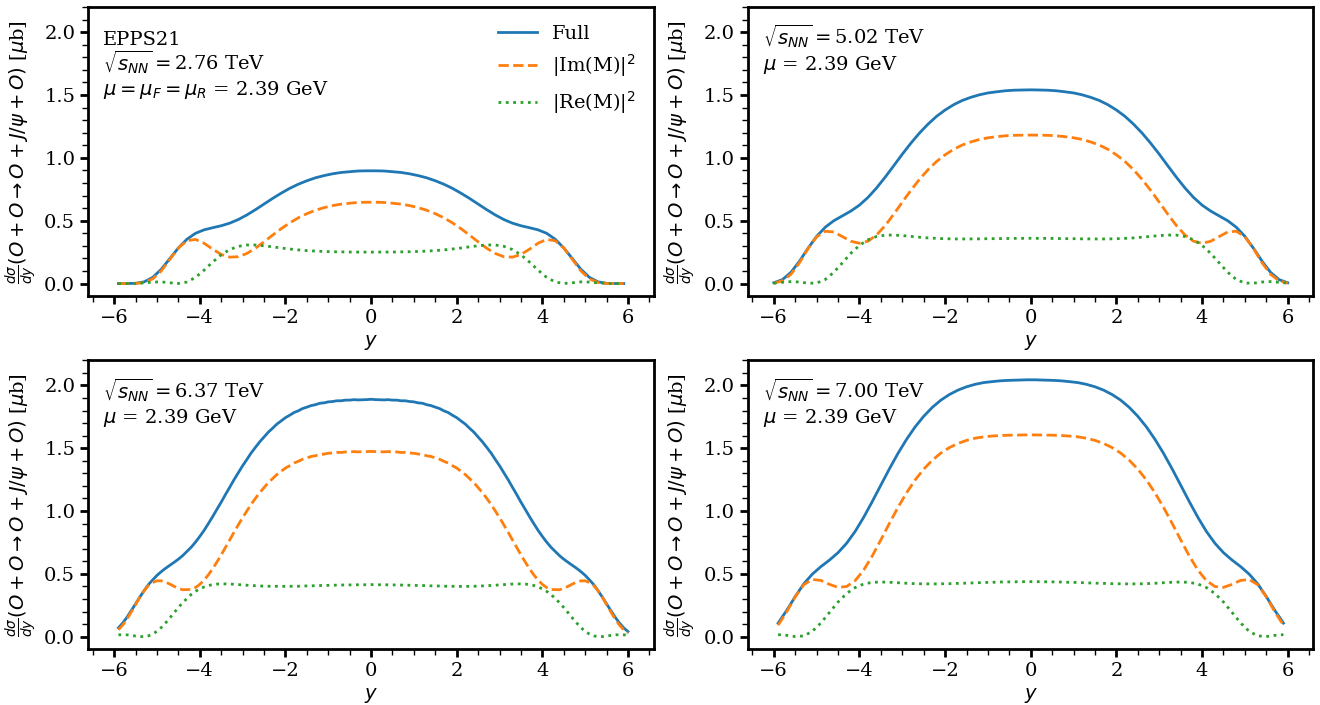}\vspace{-0.3cm}
\caption{The contributions of the imaginary (dashed orange curve) and real (green dotted curve) parts of the $\gamma +A \to J/\psi+A$ amplitude to the $d\sigma({\rm O}+{\rm O} \to {\rm O}+J/\psi+{\rm O})/dy$ cross section of coherent $J/\psi$ photoproduction in O-O UPCs as a function of the rapidity $y$; the solid blue curve is the full result. The calculation uses the EPPS21 nPDFs at $\mu = 2.39$~GeV. The different panels correspond to $\sqrt{s_{ NN}}=2.76$, 5.02, 6.37 and 7 TeV.}
    \label{fig:OO-oI}
\end{figure*} 

\begin{figure*}[!ht]
    \centering
    \includegraphics[width=1.0\textwidth]{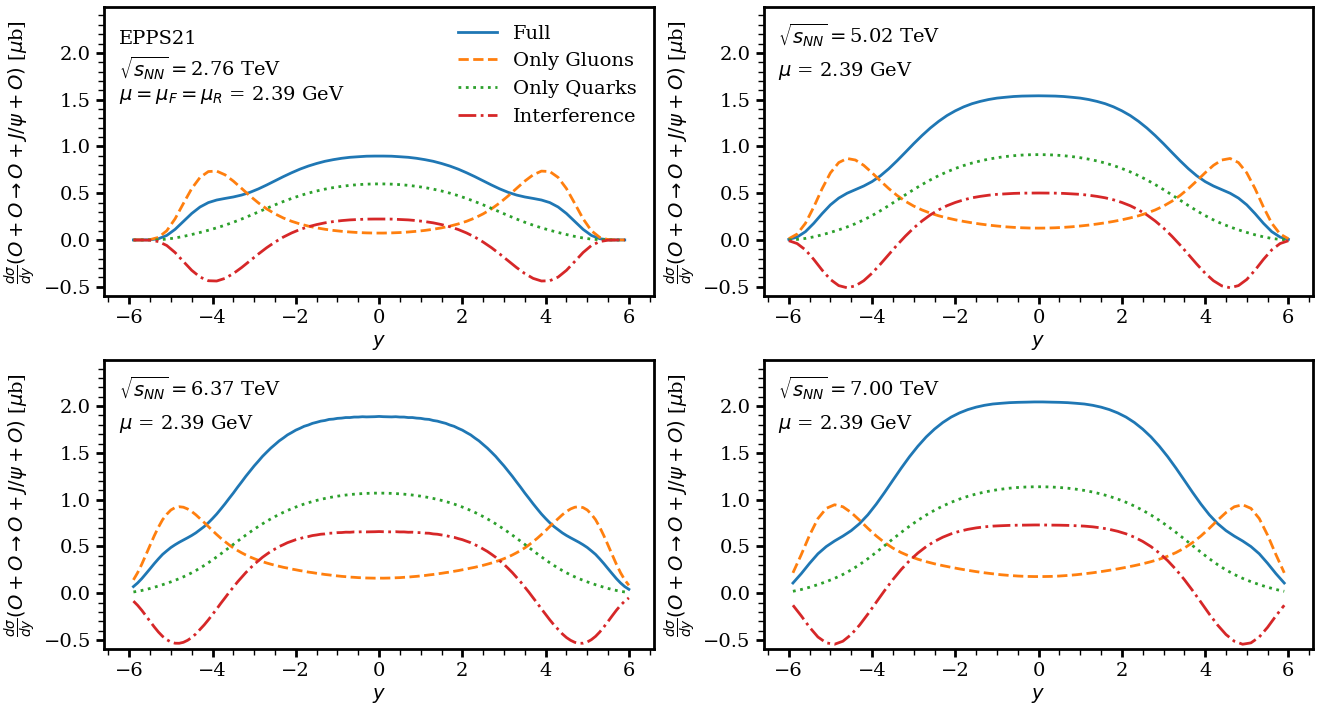}
 \caption{The breakdown of the NLO pQCD predictions for the $d\sigma({\rm O}+{\rm O} \to {\rm O}+J/\psi+{\rm O})/dy$ cross section of coherent $J/\psi$ photoproduction in O-O UPCs as a function of the rapidity $y$ into the contribution of different parton channels: gluon (dashed orange curve), quark (green dotted curve), and their interference (red dash-dot curve); the solid blue curve is the full result. The calculation uses the EPPS21 nPDFs at $\mu = 2.39$~GeV. The different panels correspond to $\sqrt{s_{ NN}}=2.76$, 5.02, 6.37 and 7~TeV.}
    \label{fig:OO-oG}
\end{figure*}

Figure~\ref{fig:OO-Wcomp} shows the separate contributions of the two terms to $d\sigma({\rm O}+{\rm O} \to {\rm O}+J/\psi+{\rm O})/dy$ in Eq.~(\ref{XS_plus_minus}), labeled ``$W^{+}$" (dashed orange) and ``$W^{-}$" (dotted green), along with their sum labeled ``Full" (solid blue). The calculation is carried out using the EPPS21 nPDFs at $\mu=2.39$~GeV, which is the optimal scale in the Pb-Pb case. The results are qualitatively similar to those for the Pb-Pb collision system~\cite{Eskola:2022vpi}. Looking only at the $W^+$ contribution, we observe a small bump at backward rapidities caused by the interplay of the large photon flux with the increasing photoproduction cross section and the integral of the nuclear form factor squared. This increase in the differential cross section is momentarily halted and then decreases as one moves from $y \approx -4$ to $y \approx -2$ (i.e. at Run~1 $\sqrt{s_{NN}}$ and slightly differently for the other energies). Then the growth of the photoproduction cross section forces an increase of the absolute magnitude of the UPC cross section until around $y \approx 2$, when the decrease in the photon flux eventually forces the cross section to zero. One can see that this holds for all four energies and we have checked that the results are qualitatively similar for all the three nPDF sets studied here.

Figure~\ref{fig:OO-oI} quantifies the contributions of the imaginary and real parts of the $\gamma +A\to J/\psi+A$ amplitude to the $d\sigma({\rm O}+{\rm O} \to {\rm O}+J/\psi+{\rm O})/dy$ UPC cross section: the dashed orange curve gives the result, when only the imaginary part is included, the dotted green curve shows the result, when only the real part is included, and the solid blue curve is their sum. One can see from the figure that with increasing $\sqrt{s_{NN}}$, the imaginary part becomes more important at central rapidity and, when moving from 2.76~TeV to 6.37~TeV, the dip in the imaginary part at around $y\pm 3$ seen at $\sqrt{s_{ NN}} = 2.76$~TeV actually rises above the real part, i.e., the imaginary part becomes the dominant contribution at all values of rapidity. Qualitatively, the results are the same for the other two nPDF sets nNNPDF3.0 and nCTEQ15WZSIH.

Finally, in Fig.~\ref{fig:OO-oG} we show the separate contributions of different parton channels to the UPC cross section. The dashed orange curve gives the gluon contribution, i.e., it corresponds to the situation when the contribution of quarks is neglected, the dotted green line gives the quark contribution, the red dash dotted curve is the interference term between the gluon and quark contributions, and the solid blue curve is the complete result. The calculation corresponds to the EPPS21 nPDFs and $\mu=2.39$~GeV. One can see from the figure that at all four considered values of $\sqrt{s_{ NN}}$, the UPC cross section at central rapidities is dominated by the quark contribution, while the gluons begin to dominate at forward and backward rapidities. We have checked that this trend also persists for the nNNPDF3.0 and nCTEQ15WZSIH nPDFs.

Lastly, a few words about the feasibility of measurements of this process in O-O UPCs. Experimentally the $d\sigma_{J/\psi}^{coh} /dy$ rapidity differential cross section for the coherent photoproduction of $J/\psi$ in the lepton channel $l^+l^{-}$ is given by~\cite{ALICE:2013wjo}
\begin{equation}
\frac{d\sigma_{J/\psi}^{coh}}{dy} = \frac{N_{J/\psi}^{coh}}{\mathcal{E}  \Gamma_{l^+l^-} \mathcal{L}_{int} \Delta y} \,,
\end{equation}
where $N_{J/\psi}^{coh}$ is the yield, i.e., the number of observed $J/\psi$ particles, $\mathcal{E}$ is the combined acceptance and efficiency of the detector, $\Gamma_{l^+l^-}$ is the branching ratio to the desired final state $l^+l^-$, $\mathcal{L}_{int}$ is the integrated luminosity, and $\Delta y$ is the width of the rapidity interval under consideration. By considering only the central rapidity and the the muon channel with $\Gamma_{l^+l^-}=5.961$\%~\cite{ParticleDataGroup:2020ssz} and taking the values given in~\cite{ALICE:2013wjo}, $\mathcal{E}=4.57$~\%, $\Delta y = 1.8$, and $N_{J/\psi}^{coh}=250$, together with $d\sigma^{coh}_{J/\psi} /dy = 2~\mu$b from Fig.~\ref{fig:OO-Wcomp}, we can estimate the required integrated luminosity $\mathcal{L}_{int}$ to be
\begin{equation}
    \mathcal{L}_{int} \approx 25.5\times 10^3 \frac{1}{\mu b} \, .
\end{equation}
It was discussed in Ref.~\cite{Citron:2018lsq} that in the high luminosity O-O run at the LHC, the average luminosity would be $\langle \mathcal{L}_{AA} \rangle = 8.99\times 10^{30}$~cm$^{-2}$s$^{-1}$. This means that in a specialized 24-hour O-O run at ALICE, the integrated luminosity would be approximately $7.8\times 10^5$ $\mu$b$^{-1}$ resulting in approximately $7.5\times 10^3$ $J/\psi$'s making the experimental data acquisition more than feasible. Unfortunately, at the proposed short data acquisition during Run 3, one would most likely acquire only the integrated luminosity of 500~$\mu$b$^{-1}$, which means that one expects to see only five events~\cite{Citron:2018lsq}.

\subsection{Ratios of O-O and Pb-Pb UPC cross sections} \label{Subsec:RatiosOfCrossSections}

\begin{figure*}[!ht]
    \centering
    \includegraphics[width=1.0\textwidth]{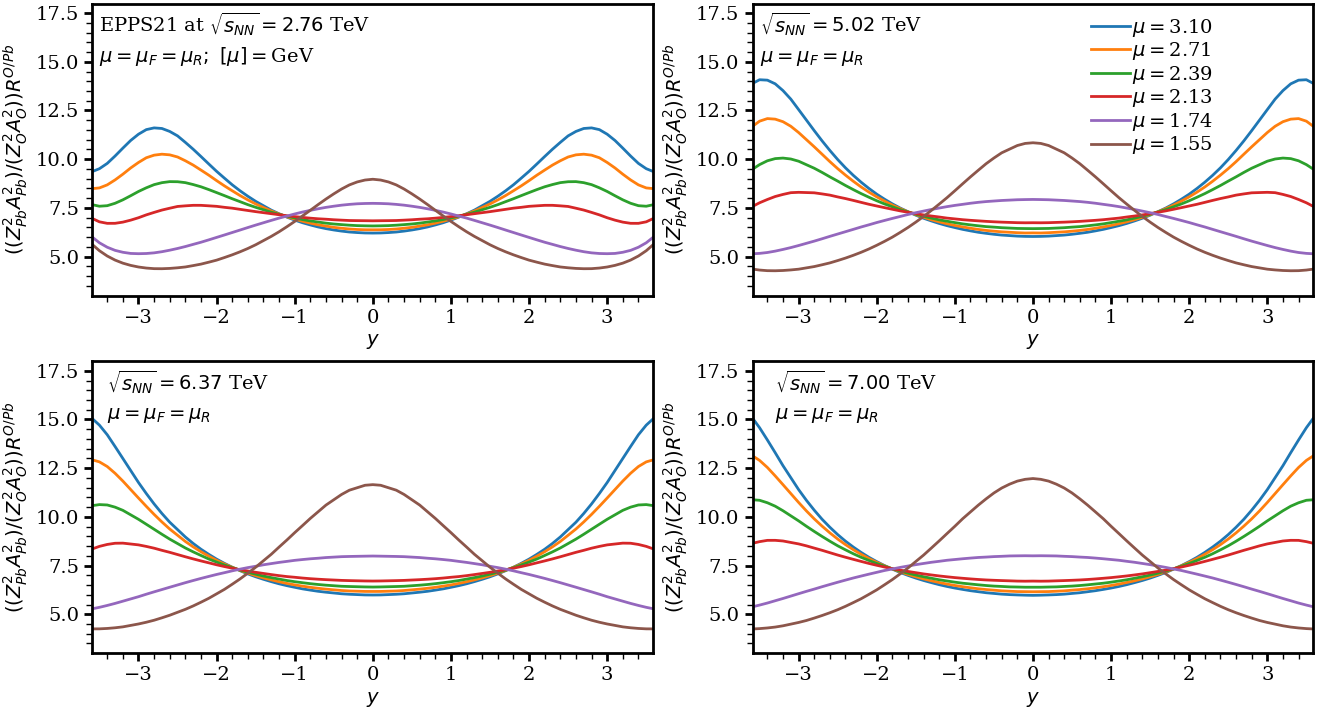}
\caption{The NLO pQCD predictions using the EPPS21 nPDFs for the scaled ratio of cross sections of $J/\psi$ photoproduction in O-O and Pb-Pb UPCs as a function of the rapidity $y$ for six different values of the scale $\mu$ at four different values of $\sqrt{s_{ NN}}$.}
    \label{fig:Rat-EPPS21-SameE} 
\end{figure*}

Our results presented above indicate that the scale dependence is considerable for both O-O and Pb-Pb collision systems. To reduce it, we examine the following scaled ratio of the O-O and Pb-Pb UPC cross section, 
\begin{equation} \label{Eq:ScaledRatio}
    R^{\rm O/Pb} = \left(\frac{208 Z_{\rm Pb}}{16 Z_{\rm O}}\right)^2 \hspace{-0.2cm} \frac{d\sigma({\rm O}+{\rm O}\rightarrow {\rm O}+J/\psi+{\rm O}) /dy}{d\sigma({\rm Pb}+{\rm Pb}\rightarrow {\rm Pb}+J/\psi+{\rm Pb})/dy} 
\end{equation}
where the factor of $[(208 Z_{\rm Pb}/(16 Z_{\rm O})]^2$ is introduced to remove the effects of the $Z^2$ scaling of the photon flux and the $A^2$ scaling of the nuclear form factor squared. Since the hard scattering part is the same for both O-O and Pb-Pb scatterings, the scale dependence, which we expect to see in this ratio, comes from the underlying nPDF sets and the different weights of the photon fluxes and the form factors, when we consider both processes at the same $\sqrt{s_{NN}}$. From a practical point of view, the O-O run will most likely be done at a different $\sqrt{s_{NN}}$, which generates an additional scale uncertainty due to the fact that the O-O process will be probed at a smaller $x$ value due to the skewness parameter $\xi$ becoming smaller.

\begin{figure*}[!ht]
    \centering
    \vspace{0.6em}
    \includegraphics[width=1.0\textwidth]{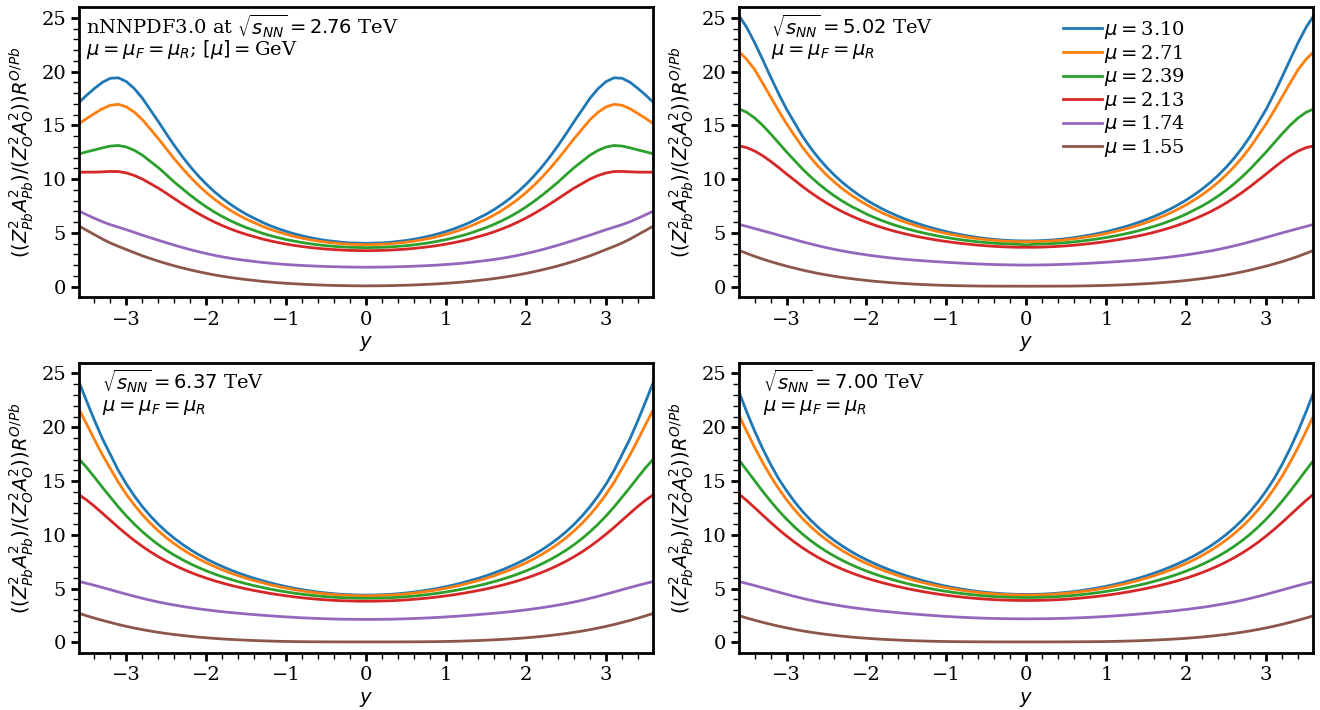}
 \caption{The same as in Fig.~\ref{fig:Rat-EPPS21-SameE}, but with the nNNPDF3.0 nPDFs.}
    \label{fig:Rat-NNPDF-SameE}
\end{figure*} 

\begin{figure*}[!ht]
    \centering
    \includegraphics[width=1.0\textwidth]{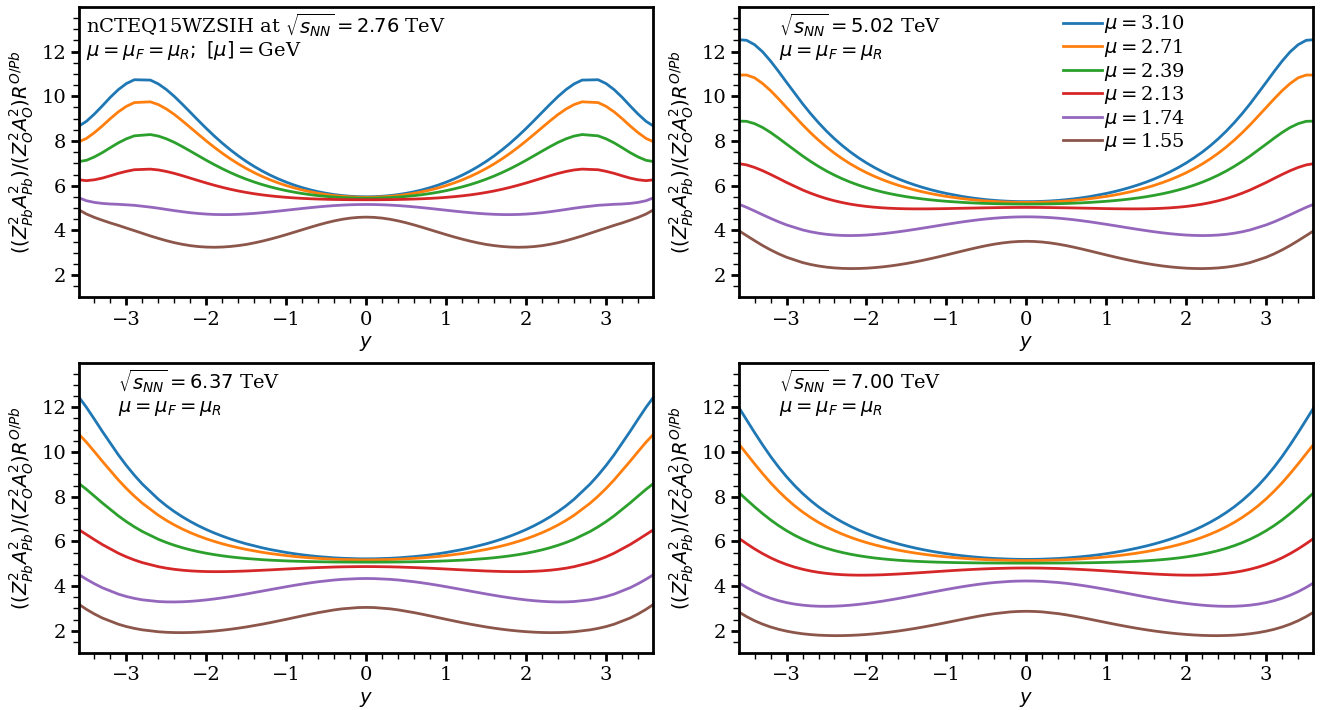}
\caption{The same as in Fig.~\ref{fig:Rat-EPPS21-SameE}, but with the nCTEQ15WZSIH  nPDFs.}
    \label{fig:Rat-CTEQ-SameE}
\end{figure*}

Figures~\ref{fig:Rat-EPPS21-SameE},~\ref{fig:Rat-NNPDF-SameE} and~\ref{fig:Rat-CTEQ-SameE} present our NLO pQCD predictions for $R^{\rm O/Pb}$ evaluated at six different values of the scale $\mu$ ranging from $\mu=1.55$~GeV to $\mu=3.1$~GeV using the EPPS21, nNNPDF3.0 and nCTEQ15WZSIH nPDFs, respectively. One can see from the figures that the relative scale uncertainty seems to be the smallest for EPPS21 and nCTEQ15WZSIH at $y \approx 0$, which then grows slightly towards backward and forward rapidities. However, in the nNNPDF3.0 case the situation is reversed due to the almost exact cancellation of the photoproduction amplitude for the O-O process at central rapidity. Moreover, depending on the energy, the EPPS21 nPDF set produces a node at $y \approx \pm 1.1$ or $y \approx \pm 1.8$, where all the scales except for the lowest $\mu = m_c$ seem to agree with each other. Such a node is missing in the results given by nNNPDF3.0 or nCTEQ15WZSIH. In addition, we would like to point out that our predictions for $R^{\rm O/Pb}$ for each nPDF set separately tend to cluster together at higher values of $\mu$.

To quantify the magnitude of the relative scale dependence, we consider the super-ratio of ratios $R^{\rm O/Pb}$ at $y=0$, which are evaluated at $\mu = M_{J/\psi}$ and $\mu = m_c$,
\begin{equation}
    R_{\text{scale}}^{\rm O/Pb} = \frac{R^{\rm O/Pb}(\mu = M_{J/\psi})}{R^{\rm O/Pb}(\mu = m_{c})} \,.
    \label{eq:R_super}
\end{equation}
The results for $R_{\text{scale}}^{\rm O/Pb}$ are presented in Table~\ref{tab:RatCentRapSameE}. One can see from the table that for all three sets of nPDFs, the scale uncertainty of $R_{\text{scale}}^{\rm O/Pb}$ is smaller by approximately a factor of 10 than that of the predictions for the individual Pb-Pb and O-O UPC cross sections (the exact size of the reduction in the scale dependence depends on the particular nPDF set and $\sqrt{s_{ NN}}$). The scale uncertainty also increases, when $\sqrt{s_{ NN}}$ is increased, since at higher energies one probes the nPDFs at progressively smaller $x$, where the scale evolution of the nPDFs is faster. 

\begin{table}[!h]
    \centering
    \caption{The ratios $R^{\rm O/Pb}(\mu = M_{J/\psi} )/R^{\rm O/Pb}(\mu = m_c )$ at $y=0$ for EPPS21, nNNPDF3.0, and nCTEQ15WZSIH nPDFs for four values of the collision energy $\sqrt{s_{NN}}$, which is taken to be the same for O-O and Pb-Pb runs.\vspace{0.5em}}
    \begin{tabular}{c|c|c|c}
         $\sqrt{s_{NN}}$ & EPPS21 & nNNPDF3.0 & nCTEQ15WZSIH  \\ \hline
         2.76 TeV        &  0.7   &   51.5    & 1.2      \\ \hline
         5.02 TeV        &  0.6   &   86.1    & 1.5      \\ \hline         
         6.37 TeV        &  0.5   &   90.6    & 1.7      \\ \hline
         7.00 TeV        &  0.5   &   91.4    & 1.8      \\ 
    \end{tabular}
    \label{tab:RatCentRapSameE}
\end{table}

\begin{figure*}[!htb]
    \centering
    \includegraphics[width=1.0\textwidth]{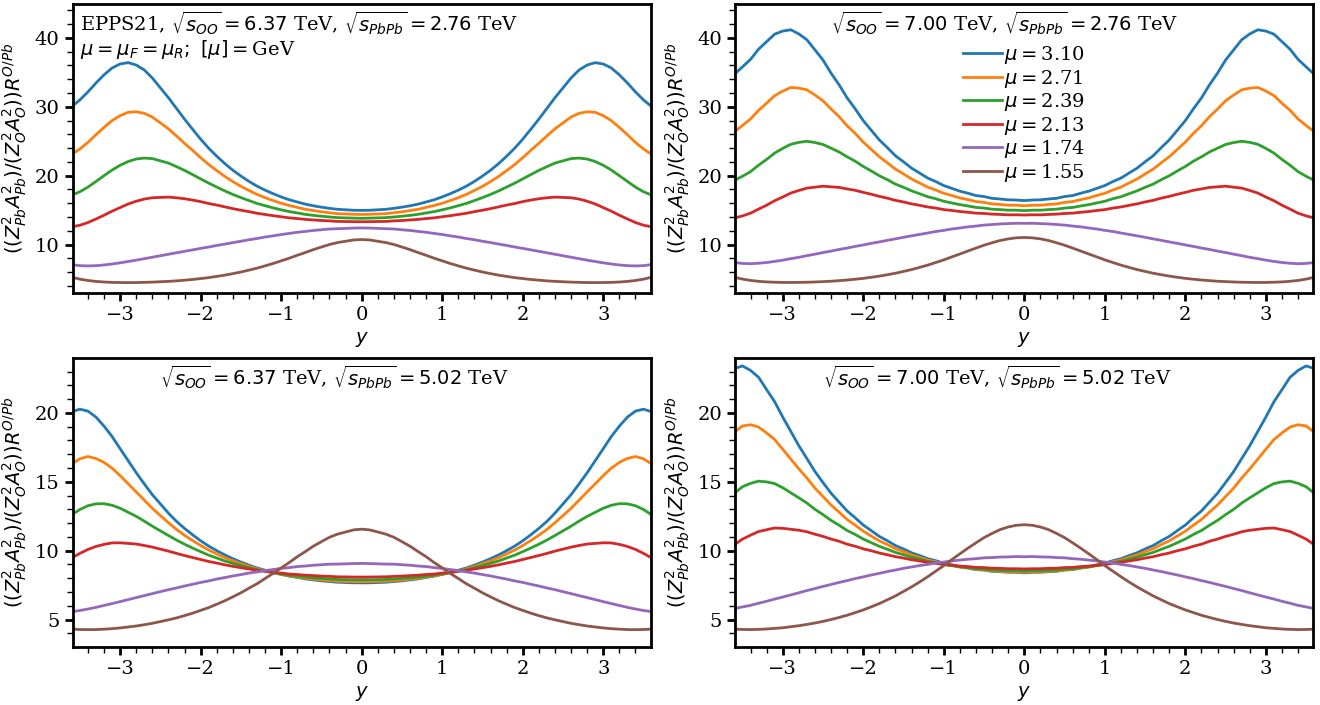}
\caption{The scaled ratio of the NLO pQCD cross sections of $J/\psi$ photoproduction in O-O and Pb-Pb UPCs as a function of the rapidity $y$ for six different values of the scale $\mu$ at non-equal values of O-O and Pb-Pb collision energies. The results are obtained with the EPPS21 nPDFs.}
\label{fig:Rat-EPPS21-DiffE}
\end{figure*}

One can see from the table that the scale uncertainty characterized by the ratio $R_{\rm scale}^{\rm O/Pb}$ of Eq.~(\ref{eq:R_super}) turns out to be very large in the case of nNNPDF3.0 nPDFs. This is an artifact of the cancellation between LO and NLO contributions to the scattering amplitude at $\mu=m_c$ that we discussed above.  If instead of $\mu=m_c=1.55$ GeV, one selects, e.g., $\mu=1.74$ GeV in the denominator of Eq.~(17), the scale uncertainty becomes dramatically reduced with $R_{\rm scale}^{\rm O/Pb} \leq 2.3$ for all four considered values of $\sqrt{s_{NN}}$, while only moderately affecting the EPPS21 and nCTEQ15WZSIH results.

\begin{figure*}[!ht]
    \centering
    \includegraphics[width=1.0\textwidth]{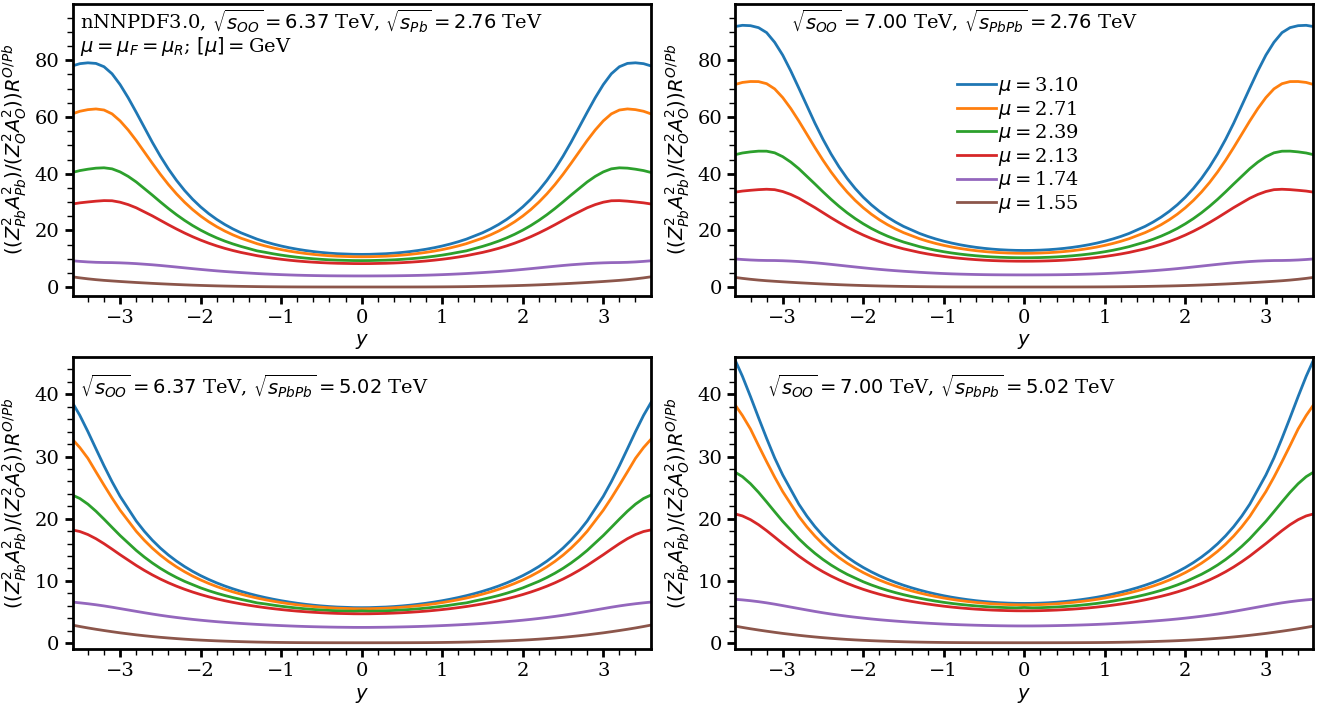}
\caption{The same as in Fig.~\ref{fig:Rat-EPPS21-DiffE}, but with the nNNPDF3.0  nPDFs.}
    \label{fig:Rat-NNPDF-DiffE}
\end{figure*}

\begin{figure*}[!ht]
    \centering
    \includegraphics[width=1.0\textwidth]{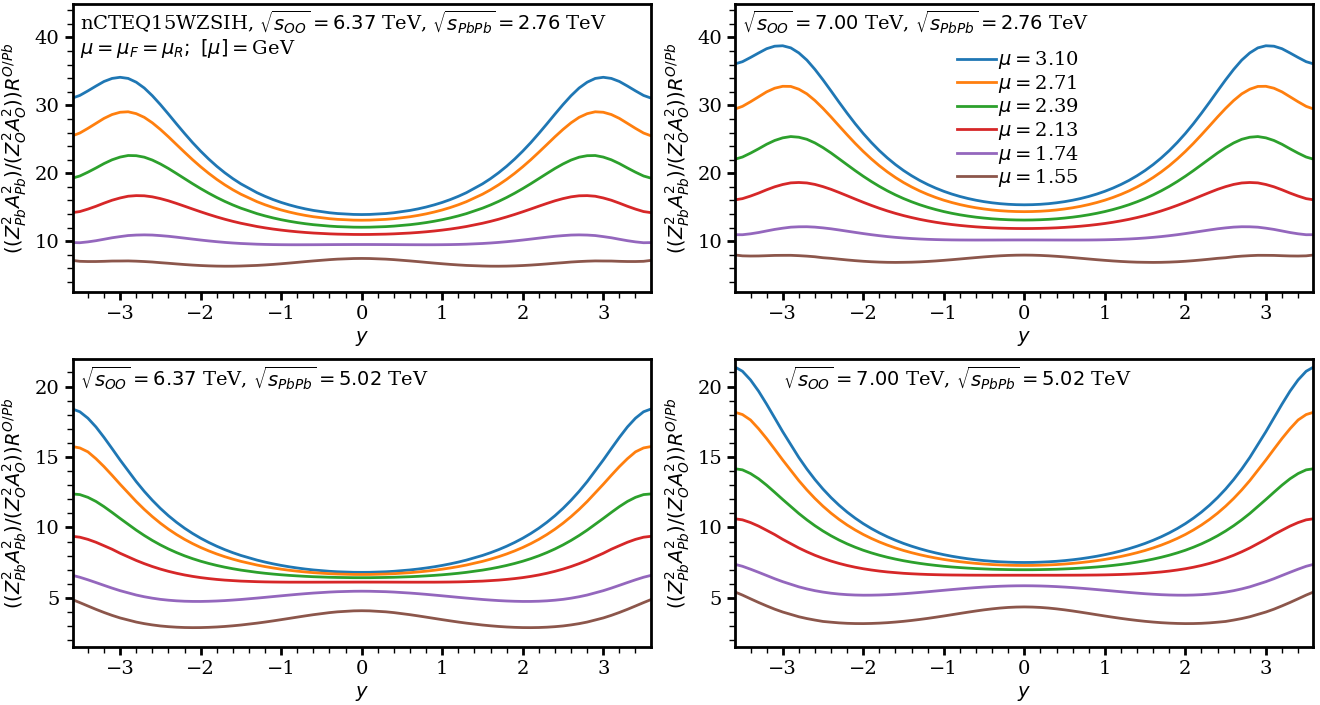}
\caption{The same as in Fig.~\ref{fig:Rat-EPPS21-DiffE}, but with the nCTEQ15WZSIH nPDFs.}
    \label{fig:Rat-CTEQ-DiffE}
\end{figure*}

To better understand the scale and energy dependence of the ratio of the O-O and Pb-Pb UPC cross sections, we consider the ratio $R^{\rm O/Pb}$, when the numerator of Eq.~(\ref{Eq:ScaledRatio}) -- the $d\sigma({{\rm O}+{\rm O}\rightarrow {\rm O}+J/\psi+{\rm O}})/dy$ cross section -- is evaluated at $\sqrt{s_{NN}}=6.37$ and 7 TeV, and the denominator of Eq.~(\ref{Eq:ScaledRatio}) -- the $d\sigma({\rm Pb}+{\rm Pb}\rightarrow {\rm Pb}+J/\psi+{\rm Pb})/dy$ cross section -- is evaluated at $\sqrt{s_{NN}}=2.76$~TeV (Run~1) and 5.02~TeV (Run~2). Our results for the EPPS21, nNNPDF3.0 and nCTEQ15WZSIH nPDF sets are presented in Figs.~\ref{fig:Rat-EPPS21-DiffE},~\ref{fig:Rat-NNPDF-DiffE} and~\ref{fig:Rat-CTEQ-DiffE}, respectively. One can see from the figures that qualitatively the scale dependence is similar for nNNPDF3.0 and nCTEQ15WZSIH, but in the case of EPPS21 for Pb-Pb UPCs at $\sqrt{s_{{NN}}}=2.76$~TeV the node disappears and the systematics of the scale dependence becomes similar for all three nPDF sets.

The general effect of taking $R_{\text{scale}}^{\rm O/Pb}$ at different energies means that the scale dependence is increased as given in Table~\ref{tab:RatCentRapDiffE}. As we take O-O consistently at a higher energy, $d\sigma /dy$ increases scale by scale, as was shown in Fig.~\ref{fig:OO_Scales}. For EPPS21 the situation is more involved since for the first two entries the scale dependence is flipped, but the magnitude of the dependence stays the same. For the last two entries -- i.e., Pb-Pb taken at $\sqrt{s_{ NN}}=5.02$~TeV -- the dependence actually gets smaller. For nNNPDF3.0 the situation is the worst: taking the ratio at different energies means that we increase the scale dependence by a factor of three at worst. For nCTEQ15WZSIH the factor is only about 1.6. However, if we disregard the lowest scale and take $\mu = 1.74$~GeV instead, the scale dependence becomes smaller for all three sets. For nNNPDF3.0 the drop is quite sizeable again: at all energies the scale dependence drops to less than a factor of 3.

\begin{table}[!h]
    \centering
\caption{The ratios $R^{\rm O/Pb}(\mu = M_{J/\psi} )/R^{\rm O/Pb}(\mu = m_c )$ at $y=0$ for the EPPS21, nNNPDF3.0, and nCTEQ15WZSIH nPDFs for different values of $\sqrt{s_{{NN}}}$ in the numerator and denominator.\vspace{0.5em} }
    \begin{tabular}{c|c|c|c}
         $\sqrt{s_{\text{OO}}}/\sqrt{s_{\text{PbPb}}}$ & EPPS21 & nNNPDF3.0 & nCTEQ15WZSIH  \\ \hline
         6.37/2.76              &  1.4   &  156.1     & 1.9       \\ \hline
         7.00/2.76              &  1.5   &  166.5     & 1.9      \\ \hline         
         6.37/5.02              &  0.7   &  104.6     & 1.7      \\ \hline
         7.00/5.02              &  0.7   &  111.7     & 1.7      \\ 
    \end{tabular}
    \label{tab:RatCentRapDiffE}
\end{table}

\begin{figure*}[!ht]
    \centering
\includegraphics[width=1.0\textwidth]{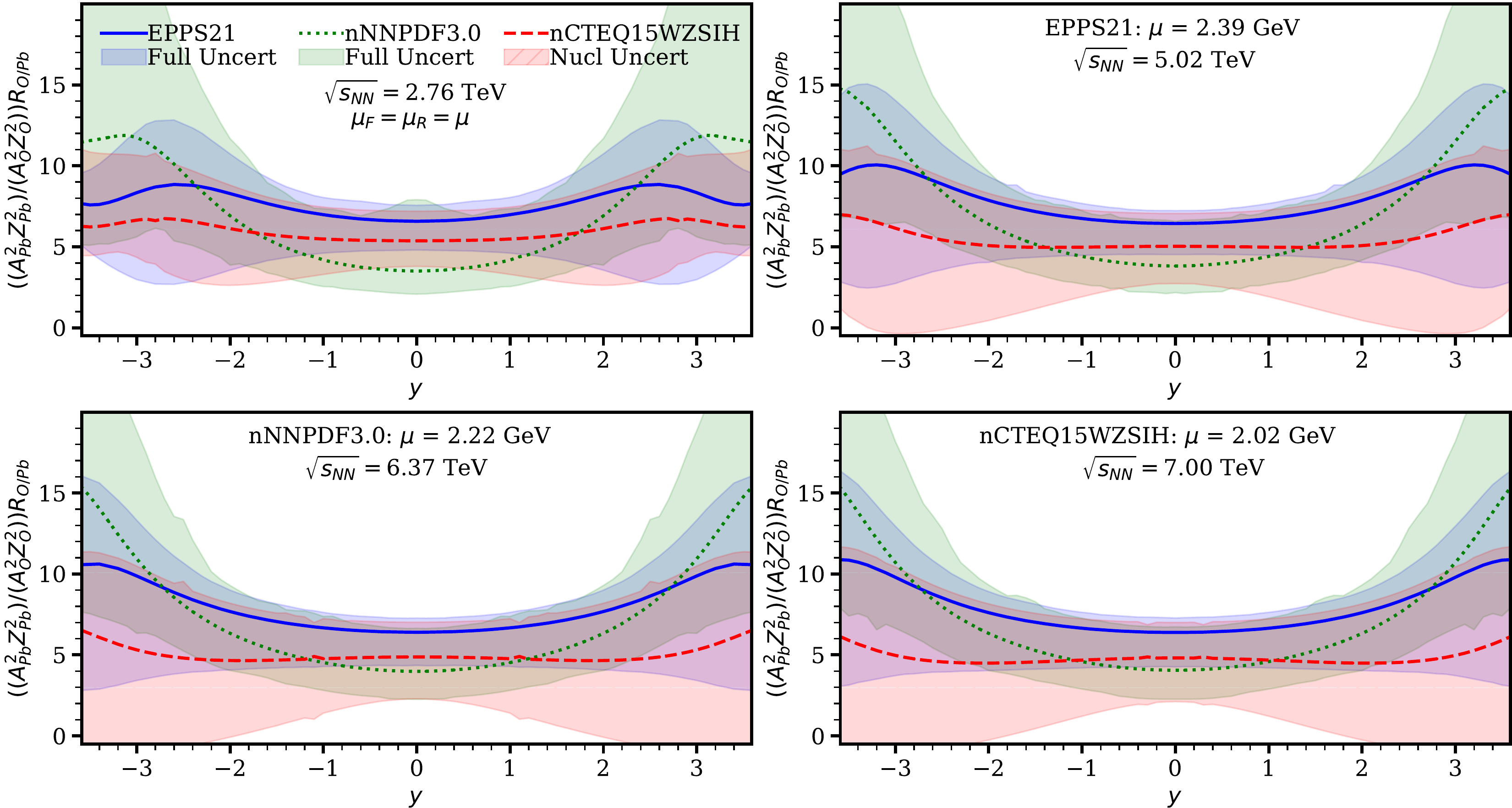}
\caption{The PDF uncertainties of NLO pQCD predictions for $R^{\rm O/Pb}$ as a function of the rapidity $y$.
The results corresponding to the central nPDF sets at the optimal scales are shown by the blue solid (EPPS21), green dotted (nNNPDF3.0), and red dashed (nCTEQ15WZSIH) curves, respectively. The corresponding uncertainties are shown by the shaded bands, see text for details. Different panels correspond to different $\sqrt{s_{NN}}$.}
    \label{fig:MainUncertSameE}
\end{figure*}

Figure~\ref{fig:MainUncertSameE} illustrates the PDF uncertainties of our NLO pQCD predictions for $R^{\rm O/Pb}$ as a function of $y$ for EPPS21, nNNPDF3.0 and nCTEQ15WZSIH nPDFs. The calculations using the central sets of the nPDFs at their corresponding optimal scales are given by the blue solid (EPPS21), green dotted (nNNPDF3.0), and red dashed (nCTEQ15WZSIH) curves. The corresponding uncertainties are given by the shaded bands. They are calculated by first finding the ratio $R^{\rm O/Pb}$ for each error set and then using the asymmetric form (see Eq.~(\ref{Eq:AsymUncert})) for EPPS21 and nCTEQ15WZSIH and the CL prescription for nNNPDF3.0. Thus, the blue and green bands give the full (free proton + nuclear) uncertainty of EPPS21 and nNNPDF3.0 nPDFs, respectively, while the hashed red band gives the nuclear uncertainty of the nCTEQ15WZSIH set. We see that the predictions agree within the PDF uncertainty bands.

One can see from the figure that as $|y|$ is increased, the uncertainty bands grow bigger for all three sets. It can be understood by noticing that at higher positive rapidities, the $W^+$ component gets probed at smaller and smaller values of $x$ (similarly for the $W^-$ component at negative rapidities). For the EPPS21 and nNNPDF3.0 sets, the band stays always at positive values, but for nCTEQ15WZSIH, the uncertainty band reaches negative values starting from $\sqrt{s_{NN}}=$ 5.02~TeV at large enough $|y|$. 

\begin{figure*}[!ht]
    \centering
 \includegraphics[width=1.0\textwidth]{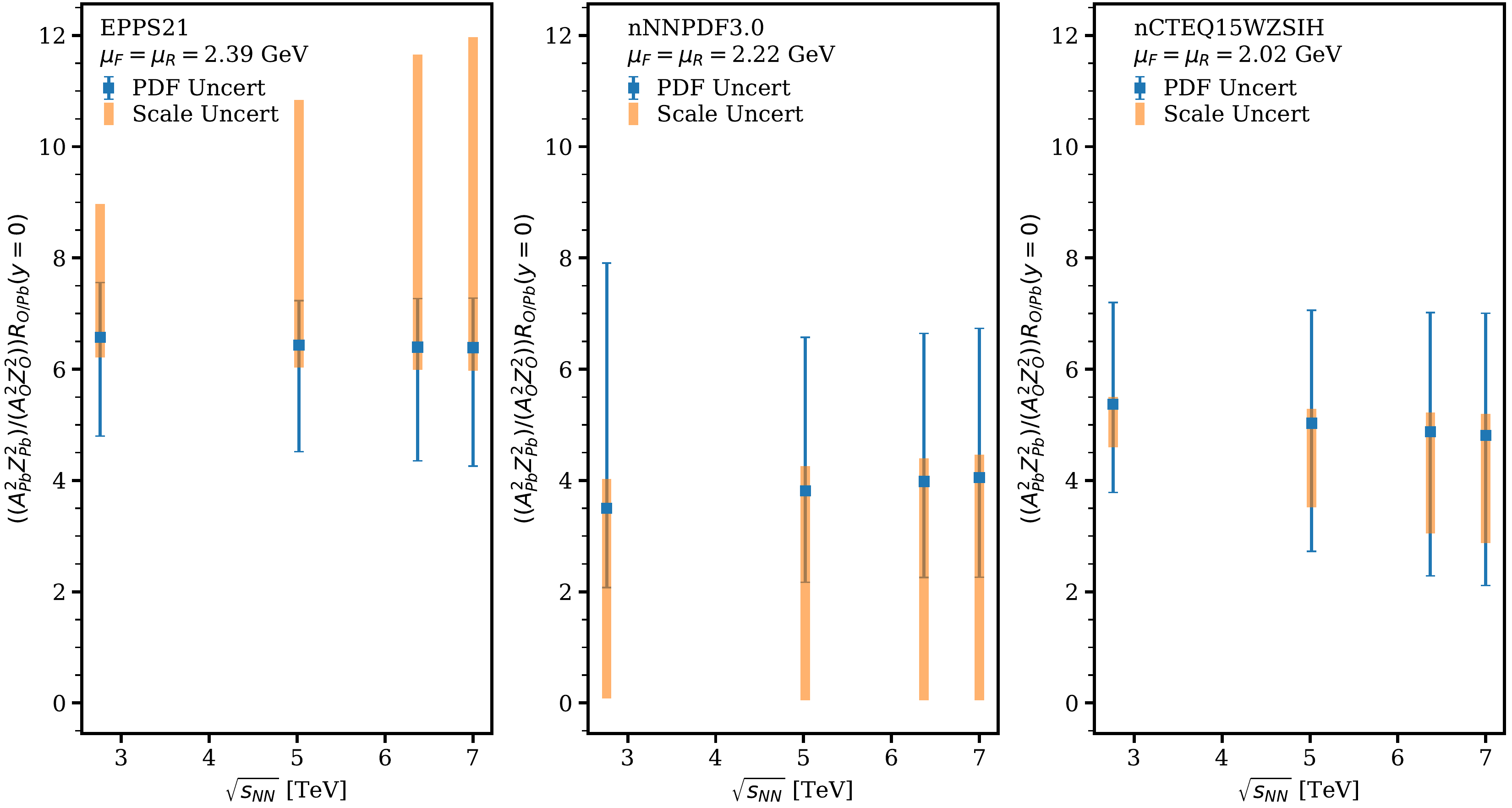}
\caption{Comparison of the PDF (thin blue) and scale (wide orange) uncertainties in the ratio of the NLO pQCD calculation of O-O to Pb-Pb rapidity differential cross section at central rapidity, $y = 0$, for three different nPDF sets: EPPS21 (left), nNNPDF3.0 (center) and nCTEQ15WZSIH (right). Here O-O and Pb-Pb are taken at same energy and all sets at their corresponding optimal scales. }
    \label{fig:MainSameE}
\end{figure*}

A comparison of the PDF and scale uncertainties in $R^{\rm O/Pb}$ at $y=0$ as a function of $\sqrt{s_{NN}}$ is shown in Fig.~\ref{fig:MainSameE}. The PDF uncertainties are calculated at the corresponding optimal scales for the EPPS21 (left), nNNPDF3.0 (middle) and nCTEQ15WZSIH (right) nuclear PDFs. The scale uncertainty represents the range between the scales $\mu = m_c$ and $\mu = M_{J/\psi}$. 
In absolute terms the EPPS21 PDF uncertainty is typically smaller than the scale uncertainty, while for nNNPDF3.0 and nCTEQ15WZSIH 
the scale and PDF uncertainties are of similar magnitudes.
The figure also shows the lack of uniformity between the uncertainties between different sets. For instance, in the EPPS21 case, the scale uncertainty dominates upwards, whereas the PDF uncertainties dominate downwards. For nNNPDF3.0, 
the situation is 
different: the scale uncertainties dominate the downwards uncertainty, while 
the PDF uncertainties are symmetric.
%
Then interestingly for nCTEQ15WZSIH -- the set with the enhanced strange quark contribution -- the scale uncertainties are smaller than the PDF uncertainties at all energies. The value of the ratio stays approximately constant as a function of $\sqrt{s_{NN}}$ for all three sets.

\begin{figure*}[!ht]
    \centering
\includegraphics[width=1.0\textwidth]{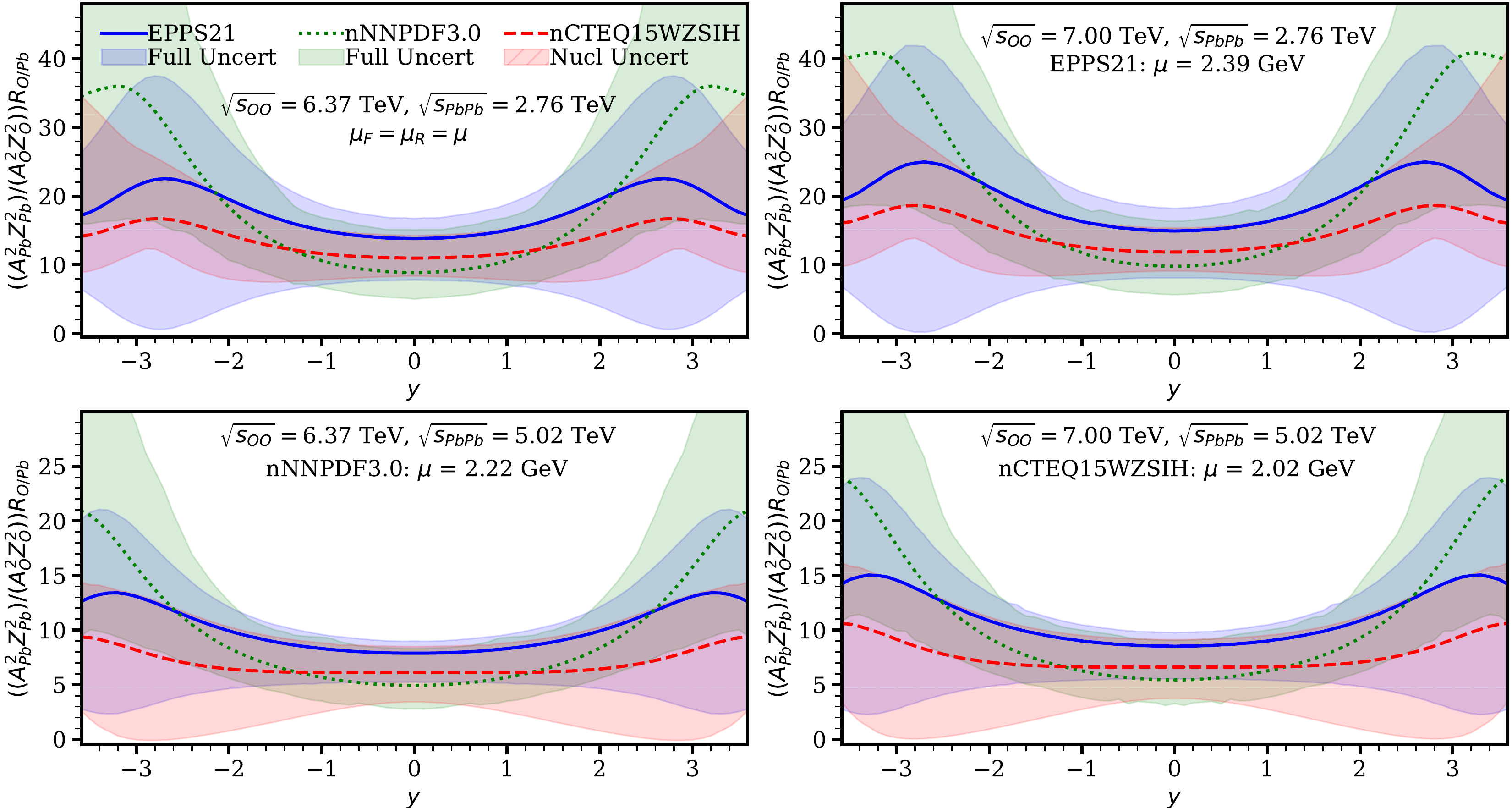}
    \caption{The scaled ratios for EPPS21 (solid blue), nNNPDF3.0 (dotted green) and nCTEQ15WZSIH (dashed red) at their optimal scales as a function of the $J/\psi$ rapidity. The blue band gives the EPPS21 uncertainty, the green band gives the nNNPDF3.0 90\% CL uncertainty and the hatched red band gives the nCTEQ15WZSIH nuclear uncertainty. In the first row Pb-Pb has been taken at Run~1 energy and in the second row at Run~2 energy. The O-O energies correspond to the two proposed energies of $6.37$~TeV (left column) and $7$~TeV (right column). }
    \label{fig:MainUncertDiffE}
\end{figure*}

Figure~\ref{fig:MainUncertDiffE} presents the nPDF uncertainties of the ratio $R^{\rm O/Pb}$ as a function of $y$, when the O-O and Pb-Pb UPC cross sections are evaluated at different collision energies (see our discussion above). The notation of the curves and shaded bands is the same as in Fig.~\ref{fig:MainUncertSameE}. A comparison with Fig.~\ref{fig:MainUncertSameE} shows that the results in the two figures are similar. In particular, at central rapidity for EPPS21 the ratio between the upper bound and the lower bound for the PDF uncertainties is about 2.2 for Pb-Pb taken at Run~1 energy and 1.8 for Pb-Pb taken at Run~2 energy, which means that the PDF uncertainty is slightly larger at all energies under consideration than the scale uncertainty (see Table~\ref{tab:RatCentRapDiffE}). For nNNPDF3.0 the same ratio is around 3
 for all energies and again the scale uncertainty is clearly the dominating one, when considering $\mu \in [m_c,M_{J/\psi}]$. If we ignore the lowest scale $\mu = m_c$, we find that the PDF uncertainty is again the larger one. For nCTEQ15WZSIH the corresponding ratios are about 1.7 and 2.5 for Run~1 and Run~2 energies, respectively.

\begin{figure*}[!ht]
    \centering
\includegraphics[width=1.00\textwidth]{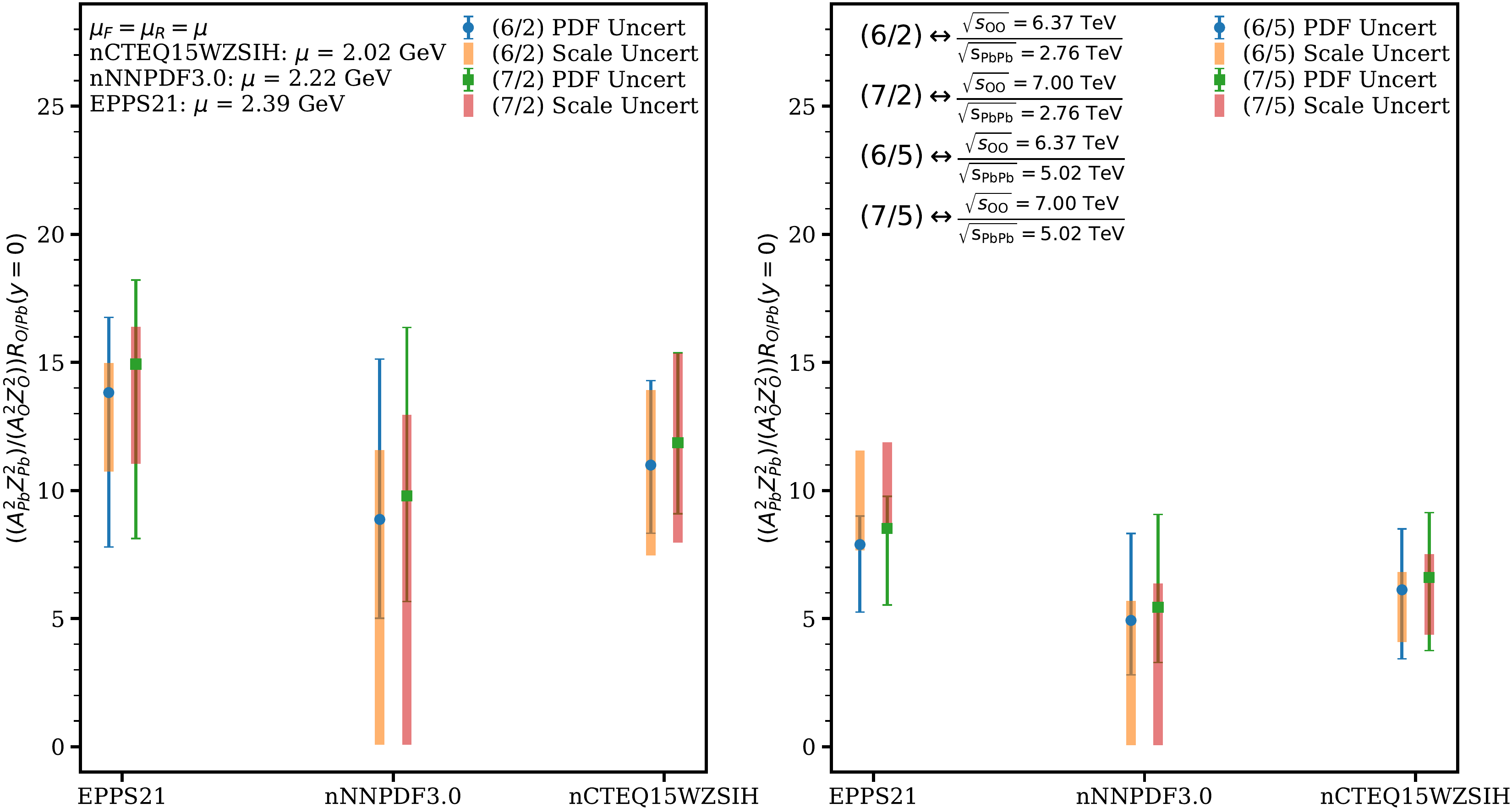}
    \caption{The scaled ratios of O-O to Pb-Pb rapidity differential cross sections for EPPS21, nNNPDF3.0 and nCTEQ15WZSIH at their corresponding optimal scales at central rapidity, $y=0$, where O-O and Pb-Pb have been taken at different $\sqrt{s_{ NN}}$ energies. In the left panel the Pb-Pb collision is taken at Run~1 energy and in the right panel at Run~2 energy.}
    \label{fig:MainDiffE}
\end{figure*}

These results are summarized in Fig.~\ref{fig:MainDiffE}, which shows the $R^{\rm O/Pb}$ ratio at $y=0$ for EPPS21, nNNPDF3.0, and nCTEQ15WZSIH for different configurations of collision energies as discussed above. The color-coded bars give the scale (wide error bars) and PDF (thin error bars with caps) uncertainties; the former are calculated using the central sets of the respective nPDF fits and the latter are evaluated at the respective values of the optimal scale $\mu$. The left and right panels correspond to the Run~1 and Run~2 energies of Pb-Pb collisions, respectively. One can see from the figure that $R^{\rm O/Pb}$ at $y=0$ and its uncertainties decrease as the Pb-Pb c.m.s. energy is increased and that $R^{\rm O/Pb}$ at $y=0$ and its uncertainties increase as the O-O c.m.s. energy is increased. One can also see that the different nPDF set predictions agree within the PDF uncertainties.

\section{\label{Sec:Conclusions}Conclusions}

This work continues our studies of $J/\psi$ photoproduction in nucleus-nucleus UPCs at the LHC within the framework of collinear factorization and NLO perturbative QCD. In particular, we update our results for this process in Pb-Pb UPCs and make predictions for the $d\sigma({\rm Pb}+{\rm Pb} \to {\rm Pb}+J/\psi+{\rm Pb})/dy$ cross section as a function of the $J/\psi$ rapidity $y$ using the state-of-the-art EPPS21,  nNNPDF3.0, and nCTEQ15WZSIH nPDF sets. Taking nuclear generalized parton distribution functions in their forward limit, where they reduce to the nPDFs, we obtain a good description of Run~1 and Run~2 LHC data on $d\sigma({\rm Pb}+{\rm Pb} \to {\rm Pb}+J/\psi+{\rm Pb})/dy$. This is achieved by choosing an ``optimal" scale for each set of nPDFs: $\mu= 2.39$~GeV for EPPS21, $\mu=2.22$~GeV for nNNPDF3.0 and $\mu=2.02$~GeV for nCTEQ15WZSIH. 

Compared to our earlier calculations using EPPS16, nNNPDF2.0, and nCTEQ15 nPDFs~\cite{Eskola:2022vpi}, we can make the following observations. The results employing the central set of the EPPS21 nPDFs are found to be similar to those with the EPPS16 nPDFs with the corresponding ``optimal" scale  $\mu=2.37$~GeV. In addition, with the EPPS21 set, the PDF uncertainties have reduced significantly. At the same time, our results with the nNNPDF3.0 nPDFs exhibit a much more regular behavior than those corresponding to the nNNPDF2.0 nPDFs and, as a result, better reproduce the data. This is due to the fact that the gluon distribution in nNNPDF3.0 grows at small $x$ much slower than that in nNNPDF2.0 nPDFs. The best description of the data at both central and forward/backward rapidities at Run~1 and Run~2 energies is achieved with the nCTEQ15WZSIH, which also performs better than the nCTEQ15 set. This is due to the strongly enhanced strange quark content at small $x$. Thus, at least at NLO pQCD, this process is a potential probe of the elusive strange quark PDFs. It should be kept in mind that, because the scale dependence is significant, the situation may still change at NNLO.

We also made detailed predictions for the ${\rm O}+{\rm O} \rightarrow {\rm O}+J/\psi+{\rm O}$ rapidity-differential UPC cross section in anticipation of the planned oxygen run at the LHC. Comparing with Pb-Pb UPCs, we observe that the shape of the rapidity distribution in the O-O case is qualitatively similar to that in Pb-Pb, but the former begins to develop a valley-like structure around $y=0$ at high enough scales $\mu \sim M_{J/\psi}$. At central rapidity, the scale dependence of our results for O-O corresponding to the EPPS21 and nCTEQ15WZSIH nPDFs is slightly smaller than that for Pb-Pb collisions, but it is still of the same order of magnitude. For nNNPDF3.0, the situation is worse: the scale uncertainty grows to the order of $10^3$ due to the nearly perfect cancellation of the sum of the LO and the NLO contributions both in the real part and the imaginary parts of the amplitude at the smallest scale of $\mu = m_c$. 

The decomposition of the O-O results into the $W^{\pm}$ components, the imaginary and the real parts, and the gluon and quark contributions did not differ significantly from the results in the Pb-Pb case~\cite{Eskola:2022vpi}. Namely, the $W^{\pm}$ contributions exhibit a two-bump structure; the imaginary part gives the dominant contribution over a larger range of $y$, while the real part cannot be neglected, especially for large values of $|y|$; the quark contribution dominates at central rapidity, but the gluons become important at backwards or forwards rapidities. Furthermore, the interplay between the gluon and the quark contributions plays an important role.

In order to reduce the significant scale and nPDF uncertainties, we have studied the ratio of the $J/\psi$ rapidity distributions in O-O and Pb-Pb UPCs at different collision energies $\sqrt{s_{NN}}$. We found that for EPPS21 and nCTEQ15WZSIH, the scale uncertainties in the ratio indeed became significantly smaller. The reduction in the scale dependence is largest at central rapidities and slightly smaller towards backward and forward rapidities both when the ratio is taken at the same value of $\sqrt{s_{NN}}$ and when taken at different values. For nNNPDF3.0, the situation is the same at central rapidity, i.e., the ratio has a smaller scale dependence when compared to the O-O case. Interestingly, and contrary to EPPS21 and nCTEQ15WZSIH, the scale dependence of the nNNPDF3.0 ratio at forward and backward rapidities becomes even smaller since the LO and NLO contributions in the O-O results no longer cancel to such an exact degree.

The PDF uncertainties for the ratios of the rapidity distribution in O-O to Pb-Pb UPCs for EPPS21
were found to be smaller 
than those for nCTEQ15WZSIH and nNNPDF3.0.
 This is a direct consequence of the tightly constrained error sets in EPPS21, whereas in nCTEQ15WZSIH and nNNPDF3.0, there is more variation. The comparison of the PDF and scale uncertainties for the ratios taken at the same energy shows that the scale uncertainty is the dominant one for the EPPS21, while for nCTEQ15WZSIH the situation is reversed. 
 For nNNPDF3.0, the two uncertainties are similar.
 For the ratios taken at different energies, the PDF uncertainties are of the same magnitude for EPPS21 and nCTEQ15WZSIH
 and somewhat larger for nNNPDF3.0.

Our analysis demonstrates that the large scale uncertainty of our NLO pQCD results can be tamed through suitably considered ratios of rapidity differential cross sections. In future work, it would be instructive to extend our analysis to $J/\psi$ photoproduction in p-Pb and p-O asymmetric UPCs and also to photoproduction of $\Upsilon$ mesons in nucleus-nucleus UPCs. In addition, our framework could be improved through a more detailed GPD modeling~\cite{Freund:2002qf} and the inclusion of the non-relativistic QCD corrections to the charmonium wave function~\cite{Zha:2018ywo,vonWeizsacker:1934nji,Jackson:1998nia,Mantysaari:2021ryb}.

\vspace{0.5cm}

\textit{Acknowledgements} 
We acknowledge the helpful discussions with I.~Helenius, H.~M\"antysaari and J.~Penttala. We acknowledge the financial support from the Magnus Ehrnrooth foundation (T.L.), the Academy of Finland Projects No. 308301 (H.P.) and No. 330448 (K.J.E.). This research was funded as a part of the Center of Excellence in Quark Matter of the Academy of Finland (Projects No. 346325 and No. 346326). This research is part of the European Research Council Project No. ERC-2018-ADG-835105 YoctoLHC.

\bibliographystyle{JHEP-2modlong.bst}
\bibliography{biblio}

\end{document}